\documentclass[twocolumn]{svjour3}

\makeatletter
\def\@seccntformat#1{\@ifundefined{#1@cntformat}%
   {\csname the#1\endcsname\quad}  
   {\csname #1@cntformat\endcsname}
}
\let\oldappendix\appendix 
\renewcommand\appendix{%
    \oldappendix
    \newcommand{\section@cntformat}{\appendixname~\thesection\quad}
}
\makeatother

\usepackage{amsmath,amssymb, amsthm}
\usepackage{stmaryrd}
\usepackage[utf8]{inputenc}
\usepackage[T1]{fontenc}
\usepackage[english]{babel}
\usepackage{indentfirst}
\usepackage{algorithmicx,xpatch}
\usepackage{algorithm}
\usepackage{algpseudocode}
\usepackage{pgf-pie}
\usepackage{listings}
\usepackage{tikz}
\usepackage{setspace}
\usepackage{multirow}
\usepackage{hyperref}
\usepackage[noend]{algcompatible}
\usepackage{url}
\usetikzlibrary{arrows,shapes,positioning,shadows,trees}
\usepackage{caption}
\usepackage{moresize}
\usepackage{pgf}
\usepackage{ipa}
\usepackage{tabularx}
\usepackage{lscape}
\usepackage{moreverb}
\usepackage{listings}

\algnewcommand\algorithmicreturn{\textbf{return}\;}
\algnewcommand\RETURN{\State \algorithmicreturn}%

\newtheorem{exemple}{Example}

\title{Faster Multiplication over $\mathbb F_2[X]$ using \texttt{AVX512} instruction set and \texttt{VPCLMULQDQ} instruction}
\author{Jean-Marc Robert  \and
           Pascal V\'eron
           \noindent\\\\
           {\rm This  is  a  pre-print  of  an  article  published  in  ``Journal  of Cryptographic Engineering''. The final authenticated version is available online at:
\texttt{https://doi.org/10.1007/s13389-021-00278-3}}
}

\authorrunning{J.-M. Robert, P. V\'eron} 

\institute{J.-M. Robert, P. V\'eron \at
              Institut de
  Math\'ematiques de Toulon\\ Universit\'e de Toulon, France \\
              \email{jean-marc.robert@univ-tln.fr}\\
              \email{veron@univ-tln.fr} \\
}

\begin{document}
\maketitle

\begin{abstract}
Code based cryptography is one of the main proposition for the post-quantum cryptographic context, and several protocols of this kind have been submitted on the NIST platform. Among them, BIKE and HQC 
are part of the five alternate candidates selected in the third round of the NIST standardization process in the KEM category. These two schemes
make use of multiplication of large polynomials over binary rings, and due to the polynomial size (from 10000 to 60000 bits), this operation is one of the costliest during key generation, encapsulation or decapsulation mechanisms. In BIKE-2, there is also a polynomial inversion which is time consuming and this problem has been addressed in \cite{DruckerGK20}.
In this work, we revisit the different existing constant-time algorithms for arbitrary polynomial multiplication. We explore the different Karatsuba and Toom-Cook constructions in order to determine the best combinations for each polynomial degree range, in the context of \texttt{AVX2} and \texttt{AVX512} instruction sets. This leads to different kernels and constructions in each case. In particular, in the context of \texttt{AVX512}, we use the \texttt{VPCLMULQDQ} instruction, which is a vectorized binary polynomial multiplication instruction. This instruction deals with up to four polynomial (of degree up to 63) multiplications, that is four operand pairs of 64-bit words with 128-bit word storing each results, the four results being stored in one single 512-bit word. This allows to divide by roughly 3 the retired instruction number of the operation in comparison with the \texttt{AVX2} instruction set implementations, while the speedup is up to 39\% in terms of processor clock cycles. These results are different than the ones estimated in \cite{NDruckerGK18}.  To illustrate the benefit of the new \texttt{VPCLMULQDQ} instruction, we used the HQC code to evaluate our approaches. 
When implemented in the HQC protocol, for the security levels 128, 192 and 256,  
our approaches provide up to 12\% speedup, for key pair generation.
\end{abstract}

\section*{Keywords}
 Finite field multiplication, Karatsuba, Toom-Cook, post-quantum cryptography, code based cryptography, \texttt{AVX2}, \texttt{AVX512}, \texttt{VPCLMULQDQ}.

\section{Introduction}
\label{sec:introduction}

In 2017, the NIST launched a consultation dealing with the so-called "Post-Quantum Cryptography" (PQC) \cite{nistPQC}, leading to think that a practical quantum computer might appear in the next two or three decades. Among the candidates known to resist against quantum computers,  several submissions on the NIST platform are  code based protocols. The public key cryptosystem of Mc Eliece marked the beginning of code based cryptography \cite{McEliece78}. The security of most of the code based protocols relies on a decision problem which can be stated without using the terminology of coding theory: the SD (Syndrome Decoding) problem.
\\\\
\begin{tabular}{lcp{6cm}}
   \bf Input &:&  $H$ a $(k,n)$ matrix over $\mathbb{F}_2$, $s\in\mathbb{F}_2^k$ a column vector, $p$ an integer.\\
    \bf Question &:& Is there a column vector $e\in\mathbb{F}_2^n$, with at most $p$ non-zero coordinates, such that $He=s$~? 
\end{tabular}
\\\\
Although this problem is NP-complete \cite{BerlekampMT78}, in practice, the efficiency of the probabilistic algorithms devoted to solve the SD problem \cite{BaldiBCPS19} has as a consequence that code based cryptography usually suffers from huge keys. Numerous strategies have been deployed to obtain a compact representation of the key. Among them, the use of double circulant codes \cite{gaborit05} leads to secure protocols with short keys.
\begin{definition}
An $n\times n$ matrix is called a circulant matrix if each row is obtained from the previous one by a cyclic shift over one position to the right. 
$$A=\begin{pmatrix}
a_0&a_1&\dots&a_{n-2} & a_{n-1}\\
a_{n-1}&a_{0}&\dots&a_{n-3} & a_{n-2}\\
\vdots &\vdots & &\vdots & \vdots\\
a_1 & a_2 &\dots & a_{n-1}&a_0
\end{pmatrix}\,.$$
\end{definition}

In the sequel, we use some coding theory terminology. The reader may refer to \cite[chapter 1]{macwilliams1977theory} for more information on coding theory.

\begin{definition}
Let $k,n \in \mathbb N$, an $(n,k)$ linear code $\mathcal C$ over $\mathbb F_2$ is a $k$ dimensional subspace of $\mathbb F_2^n$.
\end{definition}

\begin{definition}
A parity check matrix of an $(n,k)$ linear code $\mathcal C$ is an $(n-k)\times n$ matrix $H$ over $\mathbb F_2$ such that $H{~}^tc = 0$ iff $c\in \mathcal C$.
\end{definition}

\begin{definition}
A $(2n,n)$ double circulant code $\mathcal C$ is a linear code such that~: $$\begin{array}{c}(c_0,c_1,\dots,c_{n-2},c_{n-1},c_n,c_{n+1},\dots,c_{2n-2},c_{2n-1})\in\mathcal C\\
{\Downarrow} 
\\
(c_{n-1},c_0,\dots,c_{n-3},c_{n-2},c_{2n-1},c_{n},\dots,c_{2n-3},c_{2n-2})\in\mathcal C\,.
\end{array}$$
A parity check matrix $H$ of the $(2n,n)$ double circulant code has the following form~:
$$H=\bigg ( A\ \bigg |\ M\bigg ),$$
where $A$ and $M$ are two $n\times n$ circulant matrices. 
\end{definition}

A parity check matrix of a $(2n,n)$ double circulant code can be stored, in a compact way, using only its first row.
There is no general complexity result for the SD problem where $H$ is the parity check matrix of a random double circulant code. However, in practice, up to a small factor, the best attacks against the SD problem in this case are the same as those for random binary codes. Indeed, according to \cite{GaboritG07}, when $n$ is prime and $2$ is a primitive root of $\mathbb Z/n\mathbb Z$, almost all random
double circulant codes lie on the Gilbert-Varshamov
bound. As a result, the SD problem is considered hard by the cryptographic community for double circulant codes.
\\

\begin{sloppypar}
Let $y=(y_0,\dots,y_{2n-1})\in\mathbb{F}_2^{2n}$ a column vector and let us define $y^{(1)}(X)=y_0+y_1X+\cdots+y_{n-1}X^{n-1}$ and 
$y^{(2)}(X)=y_n+y_{n+1}X+\cdots+y_{2n-1}X^{2n-1}$.
Given that the algebra of $n\times n$ circulant matrices over $\mathbb{F}_2$ is isomorphic to the algebra of polynomials in the ring $\mathbb{F}_2[X]/(X^n-1)$, through the mapping $\psi$ such that $\psi(A)=a_0+a_1X+a_2X^2+\cdots+a_{n-1}X^{n-1}$, then the product $Hy$ boils down to the computation of two polynomial multiplications, namely:
\end{sloppypar}
$$\psi(A)\times y^{(1)}(X)\pmod{X^n-1}$$
\mbox{and} 
$$\psi(M)\times y^{(2)}(X)\pmod{X^n-1}\,.$$
BIKE \cite{Bike19} and HQC \cite{hqc19} make use of this isomorphism which maps a matrix-vector product into a polynomial multiplication in $\mathbb{F}_2[X]/(X^n-1)$.   
Due to the polynomial size (from 10000 to 60000 bits) in both protocols, it turns out that 
this operation has an impact on key generation, key encapsulation and key decapsulation mechanisms.

For example, in the HQC submission, key generation requires one multiplication, encapsulation requires two multiplications and decapsulation requires three multiplications. These multiplications are computed over $\mathbb F_2[X]/(X^N-1)$, with $17669 \leqslant N \leqslant 57637$.

Moreover, the multiplications are performed with one sparse operand, while the other one is dense. Sparse-dense multiplications are classically implemented using convolution approaches which is an adapted version of the schoolbook approach (for example, see Aranha \emph{et al.} in \cite{AranhaBG19}).
	


The report \cite{nistFirstRoundReport} mentions that side-channel resistance is a desirable security property for NIST PQC candidates. A minimum requirement for cryptographic primitives to ensure this property is to provide constant time implementations. 
In BIKE and HQC, some secret data are represented as sparse polynomials used as operand of a multiplication by an arbitrary polynomial, in the three steps of the protocol: key generation, encryption and decryption mechanisms.
Thus, any multiplication algorithm taking advantage of the sparsity of the secret data may leak some information on it. An adversary able to exploit such source of leakage may recover information on secret data. That is why dense-dense approaches, which process the sparse operand as an arbitrary polynomial, are to be considered as mandatory.

Dense-dense multiplication over $\mathbb F_2[X]$ has been intensively studied in the past, for different applications:
\begin{itemize}
	\item schoolbook approaches (with quadratic complexity);
	\item Karatsuba-Offmann \cite{KaratsubaO62} and Toom-Cook \cite{Bodrato07} subquadratic methods, with interpolation-evaluation algorithms;
	\item Schonh\"age-Strassen \cite{SchonhageS71} and F\"urer \cite{Furer07} FFT based methods, and recent works (see Harvey \emph{et al.} in \cite{HarveyHL17,HarveyH19}) showing a quasi-linear complexity in $\mathcal O(n \log n)$ for integer multiplication.
\end{itemize}	

One reference for the dense-dense operation is the general purpose NTL library (see \cite{ntl}), which aims to provide the whole set of operations, and is based on the \texttt{gf2x} library \cite{gf2x} for the characteristic 2 operations. All the different approaches mentioned above are implemented. However, this library and the underlying \texttt{gf2x} have been designed for general purpose use and are optimized for generic operations with operand of any size.

In terms of operand size, the HQC \cite{hqc20} protocol deals with polynomials whose size is fixed as a protocol parameter. The range of sizes corresponds to the one for which the Karatsuba and Toom-Cook approaches are the best in the state-of-the-art. Indeed, in the \texttt{gf2x} library, over $\mathbb F_2[X]/(X^N-1)$, with $17669 \leqslant N \leqslant 57637$, the computations make use of Karatsuba or Toom-Cook (split by 3 or 4) approaches, while the threshold for FFT approaches is above 240000 bits (\textit{i.e.} the degree of the polynomials). These bounds are relevant with the recent results of Harvey \emph{et al.} in \cite{HarveyHL17,HarveyH19}. 

In terms of software implementation on \texttt{x86-64} platforms, until recently, the state-of-the-art was \texttt{AVX2} instruction set implementations, especially using the~\texttt{PCL\-MUL\-QDQ} instruction (see the \texttt{gf2x} library). This instruction performs a binary polynomial (of degree at most 63) multiplication over $\mathbb F_2[X]$. It returns the result stored in a 128-bit word, either an \texttt{xmm} 128-bit register or a same size memory storage location. The \texttt{AVX2} instruction deals with 256-bit registers or memory words and performs various vectorized operation on packed operands. In 256-bit words, one can store either 32 bytes, 16 16-bit words, 8 double words or 4 quadwords (whose size is 64 bits).

In 2018, Intel announced a new instruction set extension in the so-called \texttt{Icelake} processor generation, which extends the \texttt{AVX512} instruction set already available on some XEON processors. In particular, this architecture introduces a vectorized  \texttt{VPCLMULQDQ} instruction, which performs up to four polynomial \texttt{PCL\-MUL\-QDQ} multiplications, the four 128-bit results being stored in one single 512-bit word. Following this announcement, Drucker \emph{et al.} \cite{NDruckerGK18} proposed a software implementation (for polynomials up to degree $2^{16}-1$) using this instruction set, but could only experiment  simulations or adapted versions of their software implementations, since no platforms and no \texttt{Icelake} processors were available at this time. However, they claimed up to 50\% lowering of retired instructions and predicted the same drop in terms of processor clock cycle number execution. 
Their implementations consist in three core flows that perform schoolbook multiplication:
\begin{enumerate}
    \item a $4\times 4$ quadwords (64 bits) multiplication, written in \texttt{AVX},  using \texttt{xmm} registers and the \verb+PCLMULQDQ+ instruction,
    \item a $4\times 4$ quadwords (64 bits) multiplication, written in \texttt{AVX512}, using \texttt{ymm} registers and the \verb+VPCLMULQDQ+ instruction,
    \item a $8\times 8$ quadwords (64 bits) multiplication, written in \texttt{AVX512}, using \texttt{zmm} registers and the \verb+VPCLMULQDQ+ instruction.
\end{enumerate}
They also mention Karatsuba multiplications for operand sizes above 256 bits. They provide the source code only for the second approach ($4 \times 4$ quadwords multiplication, using \texttt{ymm} registers).

\medskip
\noindent\textbf{Contributions}

In this work, we explore the Karatsuba and Toom-Cook multiplication construction and we identify the best combinations to be used depending on the polynomial degrees. As an illustration, we applied these results on the \texttt{hqc-128} and \texttt{hqc-192} multiplications of the Optimized Implementation of the HQC release,  2020/10/01 version \cite{hqc20}. In this release, the multiplication implementation make use of a 2-recursive 3-split Karatsuba. We show that using a Tom-Cook-3 approach, this provides some speedup in comparison with the initial multiplication of this release. As a consequence, this has been integrated in the last official HQC release, 2021/06/06 version.

Then, in the context of \texttt{AVX512} instruction set, now available since the \texttt{Icelake} microarchitecture processors, we propose new implementations designed for cryptographic use of polynomial multiplications over $\mathbb F_2[X]$. We show that the elementary multiplication construction has to be a schoolbook approach up to the 256 bit operand level, while in the state-of-the-art \texttt{AVX2} context, a Karatsuba multiplication is required at the threshold of 128 bit operands.

\begin{sloppypar}
We then implement tailor made vectorized subquadratic approaches (recursive Karatsuba and Toom-Cook-3) using the \texttt{AVX512} instruction set and the vectorized  \texttt{VPCLMULQDQ} instruction in order to improve the performances, in comparison with current state-of-the-art \texttt{AVX2} implementations. We compare our implementations:
\end{sloppypar}

\begin{itemize}
    \item with the \texttt{gf2x} library;
    \item with the multiplications provided or derived from the Optimized Implementation of HQC.
\end{itemize}

Drucker \emph{et al.} in \cite{NDruckerGK18} estimated that, by using the new \texttt{AVX512} instruction set and especially the new \texttt{VPC\-LMUL\-QDQ} one, the retired instruction count might be divided by two, and the clock cycle number might be lowered in the same proportion. At the time they submitted their paper, there were no actual processor available implementing the instruction set extension with \texttt{VPCLMULQDQ}. In this paper, we checked their claims. We show that while the instruction count reduction can be overtaken, in our tests, the clock cycle number is lowered by about 39\% only.

This paper is organized as follows : in Section \ref{sec:karatsuba} we present the Karatsuba multiplication over $\mathbb F_2[X]$ and our \texttt{AVX512} software implementation, the timing results and comparison with the implementations of Drucker \emph{et al.} \cite{NDruckerGK18,DruckerGK20}, the \texttt{gf2x} library and state-of-the-art \texttt{AVX2} implementation; in Section \ref{sec:toom-cook} we present the Toom-Cook multiplication over $\mathbb F_2[X]$ and our \texttt{AVX512} software implementation, the timing results and comparison with the \texttt{gf2x} library and state-of-the-art \texttt{AVX2} implementation; in Section \ref{sec:HQC} we present the performances obtained with the HQC protocol:

\begin{itemize}
    \item when using our \texttt{AVX2} multiplications in \texttt{hqc-128} and \texttt{hqc-192} in the HQC release (round 3), 2020/10/01 version;
    \item when implementing our \texttt{AVX512} multiplications in the last official HQC release, 2021/06/06 version.
\end{itemize}

Finally Section \ref{sec:conclusion} provides some concluding remarks.

The source code of all of our implementations are available at  \url{https://github.com/arithcrypto/}, in the \texttt{AVX512PolynomialMultiplication} repository.



\section{Karatsuba multiplication: algorithms and implementations}
\label{sec:karatsuba}

The Karatsuba multiplication algorithm is the first subquadratic approach, which has been presented by Karatsuba and Offmann in \cite{KaratsubaO62}. This multiplication was first applied to large integers, but can be applied to polynomials. This classical approach has been extensively studied since then, and our work relies on all those previous works. Our main contribution here is the \texttt{AVX512} software implementation of this approach, in order to speedup the runtime execution of multiplications over $\mathbb F_2[X]$.

First, we review the subquadratic Karatsuba approaches for multiplication over $\mathbb F_2[X]$. We then present our implementations and the performance results.%

\subsection{Karatsuba algorithm}
\label{ssec:KaratRec}

One wants to multiply two arbitrary polynomials of degree at most $N-1$, and the result is of degree at most $2\cdot N -2$. 
The Karatsuba complexity applied to polynomial multiplication over $\mathbb F_2[X]$ has been studied by Nègre and Robert in \cite{NegreR13}. Let $A$ and $B$ be two binary polynomials of degree at most $N-1$. These polynomials are packed into an array of 64-bit words, whose size is $\lceil N/64\rceil$. Let $t = 2^r$ with $r$ the minimum value ensuring $t \geqslant \lceil N/64\rceil$. Now, $A$ and $B$ are considered as polynomials of degree at most $64\cdot t-1$.
We reproduce the Karatsuba algorithm in Algorithm \ref{algo:karatrec} Appendix \ref{app:Karat}. From \cite{NegreR13}, the complexity of the recursive Karatsuba multiplication is : $8t^{\log_2(3)} -8t$ \texttt{XOR} between 64-bit words and $ t^{\log_2(3)}$ native 64-bit multiplication.
 We assume that this native multiplication line 2 (denoted \texttt{Mult64}) is performed using a single processor instruction: this is the case of the Intel Cores i3, i5 and i7 and above.

There are variants of these approaches, splitting the operands in all number of parts, and using an elementary multiplication which can be all sort of Karatsuba multiplication for example. These variants have been extensively studied in Weimerskirch and Paar in \cite{AWeimerskirchP06}. Algorithms  \ref{algo:karat3} and \ref{algo:karat5} present the 3-way and 5-way split Karatsuba (see Appendix \ref{app:Karat}).

In Table \ref{tab:compKarat}, we remind the complexity of recursive Karatsuba multiplication ($t$ is the size of the operands in 64-bit words).

\begin{table*}[!h]
\begin{center}
\begin{tabular}{|c|c|c|}
\hline
mult.	&	\#\texttt{pclmul}	&	\#\texttt{xor}\\
\hline
\texttt{Schoolbook}	&	$t^2$	&	$4\times t^2$	\\
\hline
\texttt{KaratRec}	&	$t^{\log_2(3)}$	&	$8\times t^{\log_2(3)}$	\\
\hline
\texttt{Karat3}	&	$6\times ((t/3)^{\log_2(3)})$	&	$48\times ((t/3)^{\log_2(3)})$	\\
\hline
\texttt{Karat5}	&	$15\times ((t/5)^{\log_2(3)})$	&	$120\times ((t/5)^{\log_2(3)})$	\\
\hline
\end{tabular}
\caption{Complexity of the Karatsuba's multiplication variants}\label{tab:compKarat}
\end{center}
\end{table*}

Apart the Schoolbook, which presents the worst complexity in terms of elementary multiplications, one can verify that the Karatsuba approaches are ordered in growing complexity for equivalent sizes.

\subsection{\texttt{AVX512} Implementation}
We propose here a little survey of the possible approaches for \texttt{AVX512} implementations. Our goal is to review the state-of-the-art (to our knowledge) and possibly propose improvements.
We started to evaluate the implementation approaches from \cite{NDruckerGK18}, which were based on the schoolbook algorithm for the 256 bit and 512 bit operand sizes. Their main claim is a reduction of 50\% of the instruction count for the kernels they presented in this paper, and while an actual processor were not available at the time of their work, they evaluated a similar improvement in terms of clock cycle number for the computation time. In a more recent work, Drucker \emph{et al.} in \cite{DruckerGK20} proposed a new approach based on the Karatsuba algorithm. In our evaluation, this last work outperforms the first of \cite{NDruckerGK18}. In this section, we briefly review the main feature of the \texttt{AVX512} instruction set, and then present the most interesting approaches for elementary multiplication for the 512 bit operand size, from \cite{DruckerGK20} and our work.

\subsubsection{\texttt{AVX512} instruction set and special features}

The \texttt{AVX512} are 512 bit extensions to the SIMD (Single Instruction Multiple Data) 256 bit \texttt{AVX} (Advanced Vector Extension) instructions for x86 instruction set architecture. This was proposed by Intel since 2013 and consists of multiple extensions. In the \texttt{AVX512} processors, in addition to the general purpose 64 bit registers, larger registers are also available in order to perform vectorized instructions. These registers are of type

\begin{itemize}
	\item \texttt{xmm} : of size 128 bits, i.e. containing two quadwords, thus denoted $\{a_1,a_0\}$, $a_0$ and $a_1$ being the quadwords in register \texttt{a};
	\item \texttt{ymm} : of size 256 bits, i.e. containing four quadwords, thus denoted $\{a_3,a_2,a_1,a_0\}$, $a_0$, $a_1$, $a_2$ and $a_3$ being the quadwords in register \texttt{a};
	\item \texttt{zmm} : of size 512 bits, i.e. containing eight quadwords, thus denoted\\ $\{a_7,a_6,a_5,a_4,a_3,a_2,a_1,a_0\}$, $a_0$, $a_1$, $a_2$ , $a_3$, $a_4$, $a_5$, $a_6$ and $a_7$ being the quadwords in register \texttt{a};
\end{itemize}

The number of registers of each type is 32.


\subsubsection{The elementary polynomial multiplication:}

Let us first remind how the \texttt{PCLMULQDQ} instruction works.

\begin{sloppypar}
The intrinsic available by including the \texttt{immintrin.h} file is :
\end{sloppypar}
\begin{verbatim}
__m128i _mm_clmulepi64_si128 (__m128i a,
          __m128i b, const int imm8)
\end{verbatim}

The quadwords $a_i$ and $b_i$ represent binary polynomials of degree at most 63. The \texttt{PCLMULQDQ} instruction returns the results in an \texttt{xmm} register, that is $a_j\times b_i$, of degree at most 126. The selection of $i$ and $j$, i.e. the corresponding quadword of the operand is made according to the value of \texttt{imm8} :
\begin{enumerate}
	\item \texttt{imm8} $= 0x00 : i=0, j=0 \rightarrow$ \texttt{PCLMULQDQ} returns $a_0\times b_0$;
	\item \texttt{imm8} $= 0x01 : i=0, j=1 \rightarrow$ \texttt{PCLMULQDQ} returns $a_1\times b_0$;
	\item \texttt{imm8} $= 0x10 : i=1, j=0 \rightarrow$ \texttt{PCLMULQDQ} returns $a_0\times b_1$;
	\item \texttt{imm8} $= 0x11 : i=1, j=1 \rightarrow$ \texttt{PCLMULQDQ} returns $a_1\times b_1$;
\end{enumerate}

The \texttt{VPCLMULQDQ} instruction now available on \texttt{Icelake} and above platforms has the following intrinsic: 

\begin{verbatim}
__m512i _mm512_clmulepi64_epi128 (__m512i a,
                   __m512i b, const int Imm8)
\end{verbatim}

This instruction computes in parallel 4 \texttt{pclmul} multiplications, i.e. carryless multiplications of binary polynomials of degree at most 63, stored in 4 quadwords in 512 bit registers, as seen above. Thus, four 128 bit results are stored in the \texttt{zmm} register as follows:

$$\{\underbrace{a_{6+j}\times b_{6+i}}_{128\ bits},\underbrace{a_{4+j}\times b_{4+i}}_{128\ bits},\underbrace{a_{2+j}\times b_{2+i}}_{128\ bits},\underbrace{a_{j}\times b_{i}}_{128\ bits}\}$$

The selection of $i$ and $j$ is made as above according to the value of \texttt{Imm8}.

\medskip
We now examine the implementation of four multiplications using this instruction set and these registers: the \texttt{mul512} version, from Drucker \emph{et al.} in \cite{DruckerGK20}, our new \texttt{karat\_1\_512} using the \texttt{mul128x4} procedure from \cite{DruckerGK20} and our new \texttt{karat\_mult\_1\_512\_SB} using a schoolbook \texttt{mul128x4} procedure, and the full schoolbook 512 bit implementation, as suggested in \cite{NDruckerGK18}. We provide the detailed source code of the \texttt{karat\_mult\_1\_512\_SB}, and corresponding explanations in Appendix \ref{app:512bitMult}, while the source code of the other approaches are available in the github repository, as mentioned in the Introduction.

We chose not to present here the schoolbook approaches of \cite{NDruckerGK18}, because in our tests, these versions are outperformed by the others. Likewise, we also implemented 256 bit kernels using schoolbook (\texttt{SB256}) and Karatsuba at the 256 bit level \texttt{Karat256} with the \texttt{AVX512} and \texttt{VPCLMULQDQ} instruction. These versions are also outperformed by the others, in particular by the \texttt{8x8} multiplication of \cite{NDruckerGK18}.

Nevertheless, we give an overview on these multiplications and their performances Appendix \ref{app:256bitMult} and Appendix \ref{app:perfProc256}, and the source code is also provided in the github repository.

\subsubsection{\texttt{mul512} version, from Drucker \emph{et al.} in \cite{DruckerGK20}}

In \cite{DruckerGK20}, Drucker \emph{et al.} present a multiplication of 1024 bit operands. This multiplication is computed as follows:

\begin{sloppypar}
\begin{itemize}
	\item The \texttt{mul1024} is a Karatsuba wrapper which calls three \texttt{mul512} multiplications, along with a classical Karatsuba reconstruction. This implementation is similar to the \texttt{AVX2} equivalent implementations except the register size.
	\item The \texttt{mul512} is a Karatsuba multiplication which splits in four parts the 512 bit operands. They use a four 512 bit word table, storing the 5 elementary xored operands in addition to the operands themselves, for a total of 9 pairs of 128 bit operands. After this step of operand preparation, the \texttt{mul512} procedure calls three times a \texttt{mul128x4} function in order to compute the 9 elementary 128 bit operand multiplications.
	\item Finally, the \texttt{mul128x4} procedure performs four 128 bit operand multiplications in parallel, using the \texttt{VPCLMULQDQ} instruction. This instruction is called three times, corresponding to the Karatsuba construction using 128 bit operands, split in two 64 bit words.
\end{itemize}
\end{sloppypar}

One may notice that the \texttt{mul512} procedure invokes 9 times the \texttt{VPCLMULQDQ} instruction, that is 36 elementary 64 bit operand multiplications, while using only $3^3 = 27$ out of them due to the Karatsuba multiplication. We refer the reader to \cite[Appendix B]{DruckerGK20}, for a complete and detailed explanation of the source code.

\subsubsection{Our new \texttt{karat\_mult\_1\_512} using the \texttt{mul128x4} procedure, from \cite{DruckerGK20}}

\begin{sloppypar}
Starting from the previous implementation of \cite{DruckerGK20}, our goal is to check the difference between the Karatsuba and the schoolbook approach at the 256 bit operand level. For this sake, we modified the \texttt{mul512} procedure into classic Karatsuba construction, which split in two parts the 512 bit operands. Now, three elementary 256 bit schoolbook multiplications are computed. These multiplications invoke one single call to the \texttt{mul128x4} procedure from \cite{DruckerGK20}. In our version, the code of the \texttt{mul128x4} procedure has been manually inlined.
\end{sloppypar}

The total number of \texttt{VPCLMULQDQ} instructions in this 512 bit multiplication is 9, the same as previously. However, this procedure now makes use of all 36 elementary 64 bit multiplication computed, while slightly simplifying the final reconstruction at the 256 bit level. This version is called \texttt{karat\_mult\_1\_512}, and its source code can be found in the github repository of the paper.

\subsubsection{Our new \texttt{karat\_mult\_1\_512\_SB} using a schoolbook \texttt{mul128x4} procedure}
\label{sec:eltKarat512}

This configuration implements the schoolbook algorithm at the 128 bit and 256 bit multiplication levels. Our goal now is to check which algorithm between Karatsuba and schoolbook is the best at the 128 bit operand level. Indeed, the instruction count is lower, while making use of one more \texttt{VPCLMULQDQ} instruction. The latency and throughput of this instruction is higher than conventional instructions. However, the vectorized version changes this by performing simultaneously four elementary 64 bit operand multiplications.

This approach now uses $3\times 4^2 = 48$ elementary 64 bit operand multiplications in total. This version is called \texttt{karat\_mult\_1\_512\_SB} and is presented in details in Appendix \ref{app:512bitMult} page \pageref{app:512bitMult}.

\subsubsection{Full schoolbook 512 bit multiplication \texttt{SB512}}
\label{sec:eltSB512}

From the description of Drucker \emph{et al.} in \cite{NDruckerGK18}, we also wrote a full schoolbook approach at the 512 bit size level.

This version now uses $4\times 4^2 = 64$ elementary 64 bit operand multiplications in total, and is called \texttt{SB512}.

\subsubsection{Instruction counts for the four 512 bit multiplication versions}

\begin{sloppypar}
The instruction count for both configurations of \texttt{mul128x4} is shown Table \ref{tab:instrCounts128-13092021}.
\end{sloppypar}

\begin{table*}[htbp]
\begin{center}
\begin{tabular}{|c|c|c|}
\hline
\multicolumn{3}{|c|}{\begin{tabular}{c}Instruction count of \texttt{mul128x4}	\\
	 128 bit size operands
\end{tabular}	}	\\
\hline
Instructions									& Drucker \emph{et al.} \cite{DruckerGK20}	&	schoolbook (this work)	\\
\hline
\texttt{\_mm512\_clmulepi64\_epi128}			&	3	&	4	\\
\hline
\texttt{XOR}									&	4	&	1	\\
\hline
\texttt{\_mm512\_mask\_xor\_epi64}				&	2	&	2	\\
\hline	
\texttt{\_mm512\_permutex\_epi64}				&	2	&	-	\\
\hline
\texttt{\_mm512\_permutexvar\_epi64}			&	1	&	1	\\
\hline
\hline
\bf \large Total						&\bf \large 	12	&\bf \large 	8	\\
\hline
\end{tabular}
\caption{Instruction count for the \texttt{mul128x4} bit multiplication versions}\label{tab:instrCounts128-13092021}
\end{center}
\end{table*}

\normalsize
\begin{sloppypar}
Instruction counts for \texttt{mul512}, \texttt{SB512}, \texttt{karat\_mult\_1\_512}  and \texttt{karat\_mult\_1\_512\_SB} are shown Table \ref{tab:instrCounts512-13092021}. For the Drucker \emph{et al.} \cite{DruckerGK20} version, we count three times the \texttt{mul128x4} instruction count plus the \texttt{mul512} instructions.
\end{sloppypar}

\begin{table*}[htbp]
\begin{center}
\begin{tabular}{|c|c|c|c|c|}
\hline
\multicolumn{5}{|c|}{\begin{tabular}{c}Instruction count	\\
	 512 bit size operands
\end{tabular}	}	\\
\hline
Instructions									&	\texttt{karat\_mult\_1\_512\_SB}	&	\texttt{karat\_mult\_1\_512}	&      \texttt{SB512} & \texttt{mul512} \cite{DruckerGK20}\\
\hline
\texttt{\_mm512\_clmulepi64\_epi128}			&	12	&	9	&   16  &	9	\\
\hline
\texttt{XOR}									&	11	&	20	&   5   &	22	\\
\hline
\texttt{\_mm512\_mask\_xor\_epi64}				&	11	&	11	&   14  &	11	\\
\hline	
\texttt{\_mm512\_permutex\_epi64}				&	-	&	-	&   -   &	6	\\
\hline
\texttt{\_mm512\_permutexvar\_epi64}			&	13	&	20	&   18  &	8	\\
\hline
\texttt{\_mm512\_permutex2var\_epi64}			&	3	&	3	&   5   &	5	\\
\hline
\texttt{\_mm512\_alignr\_epi64}					&	-	&	-	&    -   &	6	\\
\hline
\hline
\bf \large Total						&\bf \large 	50	&\bf \large 	63	&\bf\large 58 &\bf \large 	67	\\
\hline
\end{tabular}
\caption{Instruction count for the 512 bit multiplication versions}\label{tab:instrCounts512-13092021}
\end{center}
\end{table*}

\begin{sloppypar}
These instruction counts give an overview of the potential differences between the schoolbook and Karatsuba approaches at different levels. The threshold for the use of Karatsuba algorithm is at the lowest level in \texttt{AVX2} implementation. However, the vectorized \texttt{VPCLMULQDQ} instruction performs 4 multiplications at a time, with similar latency and throughput. Thus, in the context of \texttt{AVX512}, the good level for this threshold is 256 bit multiplication.
\end{sloppypar}

As expected (see Table \ref{tab:instrCounts512-13092021}), the instruction count is the lowest for the \texttt{karat\_mult\_1\_512\_SB}, corresponding to the schoolbook approach until the 256 bit operand size. The \texttt{mul512} of \cite{DruckerGK20} instruction count is the greatest among the three approaches, due to the more complex final reconstruction of each Karatsuba recursion step. 

\subsubsection{Performances}

\begin{sloppypar}
We now check the performances of all the previous approaches. The test procedure is presented Appendix \ref{app:perfProc} page \pageref{app:perfProc}. Since the \texttt{SB512} version is the slowest one, we chose to provide its performances Table \ref{tab:perfKaratRec256} page \pageref{tab:perfKaratRec256}, in Appendix \ref{app:perfProc256}.\end{sloppypar}

We present Table \ref{tab:perfKaratRec} the results for the first three approaches described above, and compare them to the classic \texttt{AVX2} implementations from the \texttt{gf2x} library \cite{gf2x} and the one from the HQC submission \cite{hqc20}. The main conclusions are as follows:

\begin{itemize}
    \item among the \texttt{AVX512} implementations, the best version is the one based on the \texttt{karat\_mult\_1\_512\_SB}, i.e. a schoolbook algorithm applied to the 128 bit and 256 bit levels, and the Karatsuba approach at the highest level (512 bit operands);
    \item in comparison with the \texttt{AVX2} implementation, this version achieves a retired instruction number divided by roughly three, and speedups (in terms of clock cycle number) are from 29.5\% to up to 39.1\% (\texttt{KaratRec}, size 131072 bit), due to the lower IPC of the \texttt{AVX512} instruction set.
\end{itemize}

\begin{table*}[htbp]
\begin{center}
\begin{tabular}{|c|c|c|c||c||c|c|}

\hline
\multicolumn{2}{|c|}{KaratRec}	&	\multicolumn{2}{c||}{\texttt{AVX2}}		&	Our impl.	&	\multicolumn{2}{c|}{\bf This work}\\
\multicolumn{2}{|c|}{}	&	\multicolumn{2}{c||}{}	&	of \cite{DruckerGK20}		&	\multicolumn{2}{c|}{\texttt{AVX512} new 512 bit op. mult.}\\
\hline
size	&			&	\texttt{gf2x}\cite{gf2x}		&			after \cite{hqc20}		&	\texttt{mul512}    	&		\texttt{karat\_mult\_1\_512}	&	\texttt{karat\_mult\_1\_512\_SB}	\\
\hline
\hline
\multirow{2}{*}{1024}	&
\# clock cycles			&	339	&	183			&	\textbf{137}&	143 &	\textbf{129}\\
\cline{2-7}
&\# instructions		&	1224	&   612		&	\textbf{254}&	228 &	\textbf{193}\\
\hline
\hline
\multirow{2}{*}{2048}	&
\# clock cycles			&	998	&	   610		&	\textbf{461}&	447 &	\textbf{423}\\
\cline{2-7}
&\# instructions		&	3892	&	   1867		&	\textbf{872}&	693 &	\textbf{581}\\
\hline
\hline
\multirow{2}{*}{4096}	&

\# clock cycles			&	2949	&   1929		&	\textbf{1425}&	1383 &	\textbf{1287}\\
\cline{2-7}
&\# instructions		&	11079	&   5684		&	\textbf{2701}&	2219 &	\textbf{1821}\\
\hline
\hline
\multirow{2}{*}{8192}	&

\# clock cycles			&	8742    &   6038		&   \textbf{4424}&	4236 &	\textbf{3977}\\
\cline{2-7}
&\# instructions		&	33182    &   17991		&   \textbf{8600}&	7051 &	\textbf{6128}\\
\hline
\hline
\multirow{2}{*}{16384}	&

\# clock cycles			&	26128   &   18327	&   \textbf{13314}&	12893 &	\textbf{12078}\\
\cline{2-7}
&\# instructions		&	100163   &   54840	&   \textbf{26099}&	21573 &	\textbf{18797}\\
\hline
\hline
\multirow{2}{*}{32768}	&

\# clock cycles			&	78889   &   59613		&   \textbf{40582}&	39038 &	\textbf{36811}\\
\cline{2-7}
&\# instructions		&	295755   &   166410		&   \textbf{79078}&	65466 &	\textbf{57525}\\
\hline
\hline
\multirow{2}{*}{65536}	&

\# clock cycles			&	226640  &   187305		&   \textbf{126386}&	121348 &	\textbf{114620}\\
\cline{2-7}
&\# instructions		&	853977  &   503014		&   \textbf{238851}&	198185 &	\textbf{174468}\\
\hline
\hline
\multirow{2}{*}{131072}	&

\# clock cycles			&	667900  &   572984		&   \textbf{382669}&	369495 &	\textbf{348982}\\
\cline{2-7}
&\# instructions		&	2516857  &   1515625	&   \textbf{719423}&	597021 &	\textbf{527205}\\
\hline
\end{tabular}
\caption{Performance comparison for Algorithm \ref{algo:karatrec}}\label{tab:perfKaratRec}
\end{center}
\end{table*}

\begin{figure}[htbp]
\begin{center}
\includegraphics[scale=0.65]{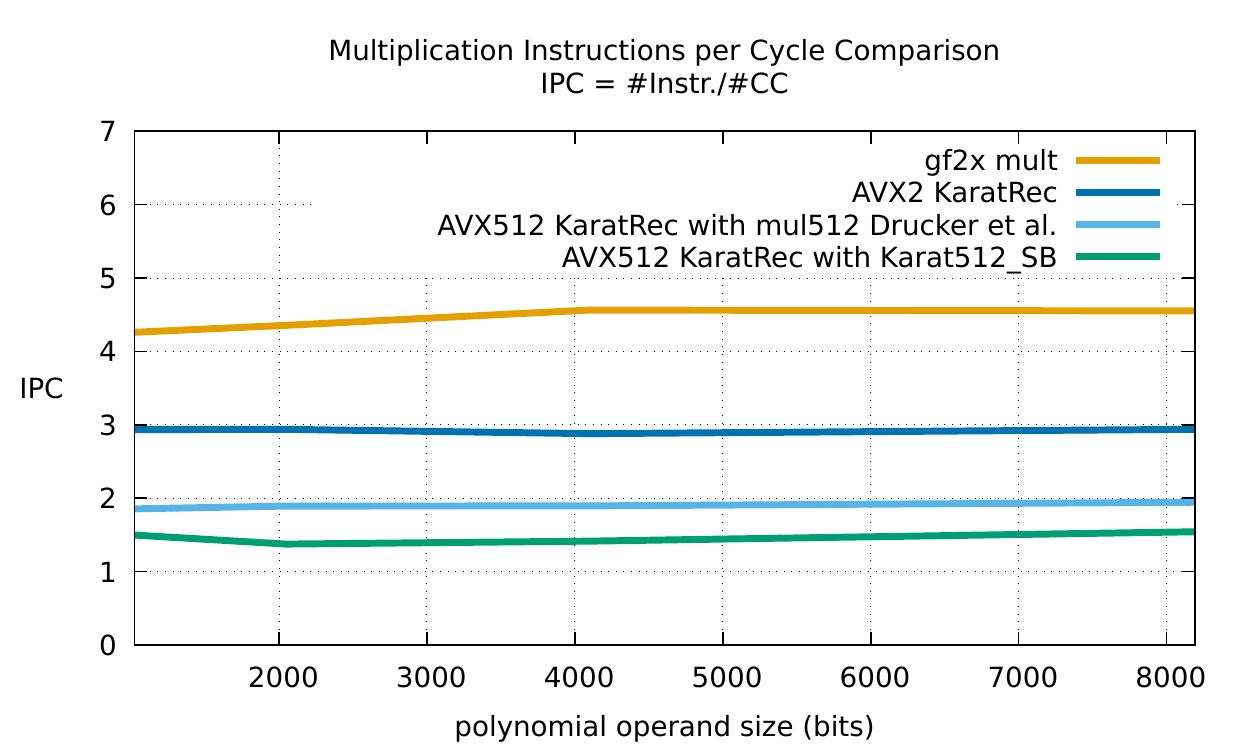}
\caption{Instructions per cycle comparison for Algorithm~\ref{algo:karatrec}}\label{fig:ilpKR_2}
\end{center}
\end{figure}

The \texttt{gf2x} library performances are slightly lower than the ones of the other implementations. The results shown Table \ref{tab:perfKaratRec} allows to evaluate the \emph{Instructions per cycles} (IPC), which is the ratio between the retired instructions and the clock cycle number. The graph in Figure \ref{fig:ilpKR_2} shows that while the IPC of the \texttt{gf2x} library multiplications is about 4.5, the IPC of the \texttt{AVX2} multiplications is nearly 3.0 and the IPC of our \texttt{AVX512} multiplications (with \texttt{karat\_mult\_1\_512\_SB}) is about 1.5. In the \texttt{gf2x} case, the software implementation makes use of \texttt{AVX2} instruction set. However, this all-purpose library includes some wrappers and tests, and especially offers the possibility to tune each operand size. This is costly, but is written with conventional instructions. This is the most likely explanation of the high IPC, while the clock cycle number is a little worse that the one of the \texttt{AVX2} software implementation.

\begin{sloppypar}
We compare also the IPC of our \texttt{AVX512} multiplication (with \texttt{karat\_mult\_1\_512\_SB}) and the one using the \texttt{mul512} of Drucker \emph{et al.} from \cite{DruckerGK20}. In this last case, the IPC is greater than the one obtained with our implementation. While the clock cycle numbers of our implementations are about 9\% lower than the ones using the \texttt{mul512}, Table \ref{tab:perfKaratRec} shows that the instruction numbers are 24\% to 33\% lower. This may be explained by the intensive use of the \texttt{\_mm512\_clmulepi64\_epi128} we have in our case. Due to this instruction high latency and throughput, the lesser instruction count does not yield such a large decrease for the clock cycle numbers. These results may also vary versus the processor version, according to the corresponding \texttt{\_mm512\_clmulepi64\_epi128} instruction latency and throughput.
\end{sloppypar}

\begin{sloppypar}
The conclusion of this comparison is that the \texttt{AVX512} multiplication implementation presents an elementary 64 bit multiplication cost relatively low in comparison with the \texttt{AVX2} situation. In this case, the non vectorized multiplication cost leads to a Karatsuba application at the 128 bit level. Insofar as the vectorized instructions equivalently divides by four the latency and the throughput of the elementary multiplication at the 64 bit level, it becomes more interesting to apply the schoolbook approach at the 128 bit level and also at the 256 bit level. In our tests, we saw that at the 512 bit level, the Karatsuba approach becomes again the best one.
\end{sloppypar}

\begin{sloppypar}
On the hardware point of view, our tests show another aspect of this \texttt{AVX512} implementation case, which differs from the initial estimations given in \cite{NDruckerGK18}, concerning the potential speedups brought by the \texttt{AVX512} instruction set and \texttt{VPCLMULQDQ} instruction. Indeed, these results make clear that, in terms of processor hardware, at the microarchitectural level, the \texttt{AVX512} instruction set has not yet been implemented with the same integration level as the one of the other instruction sets (conventional and \texttt{AVX2}), at least on our platform. The Intel documentation does not provide a lot of details \cite{IntelSD21}, however, one can assume that the hardware features are not homogeneous with the \texttt{AVX2} equivalents. This also means that future Intel processor generations might improve the IPC of the \texttt{AVX512} instruction set and potentially decrease the clock cycle number of our implementation. Indeed, between our \texttt{AVX2} and \texttt{AVX512} multiplication implementation, the retired instruction count is divided by nearly three. If the processor manufacturer improves the IPC of the \texttt{AVX512} and \texttt{AVX2} instruction sets, to get closer to the one of the conventional instruction set IPC, this means that one may observe speedups with our software implementations on future platforms.
\end{sloppypar}

\subsubsection{ \texttt{Karat3} and \texttt{Karat5} implementations}

\begin{sloppypar}
These multiplications (Algorithms \ref{algo:karat3} and \ref{algo:karat5}, Appendix \ref{app:Karat} page \pageref{app:Karat}) make use of the previous \texttt{KaratRec} multiplication (Algorithm \ref{algo:karatrec}) as elementary multiplications. The vectorized versions are again implemented using \texttt{AVX2} and \texttt{AVX512} instruction set. This leads to different sizes: while the \texttt{Karat3} multiplication has an operand size which is three times the one of the elementary multiplication (for example 512,1024, 2048... bits), the \texttt{Karat5} has an operand size which is five times the one of the elementary recursive Karatsuba multiplication. Thus, depending on the context, one may choose the most appropriate version according to the operand sizes. We present the performance results (\# clock cycles) in Table \ref{tab:perfKarat-3-5}.
\end{sloppypar}

\begin{sloppypar}
The \texttt{Karat3} or \texttt{Karat5} multiplications can also use themselves as elementary multiplications. This leads to four extra combinations for multiplication. We present the performance results (\# clock cycles) in Table \ref{tab:perfKaratCombined}. One may notice that the \texttt{Karat3(Karat5)}, i.e. a \texttt{Karat3} multiplication using a \texttt{Karat5} elementary multiplication, deals with the same operand sizes that the \texttt{Karat5(Karat3)}. However, we give the results for both versions, and observe that they present very close performances.
\end{sloppypar}

\begin{sloppypar}
In all this experimentation, we only used, as the most elementary multiplication, the \texttt{karat\_mult\_1\_512\_SB} version, presented Section \ref{sec:eltKarat512} page \pageref{sec:eltKarat512}. The platform and experimentation process are the same as previously (see Appendix \ref{app:perfProc} page \pageref{app:perfProc}).
\end{sloppypar}

\begin{table}
\begin{center}
\begin{tabular}{|c|c|c|c|c|}
\hline
	&	
\multirow{2}{*}{size}	&	\multicolumn{3}{c|}{\# clock cycles}\\
\cline{3-5}
	&&	
	\texttt{gf2x}&
		\texttt{AVX2}		&
			\texttt{AVX512}\\

\hline
\hline
\multirow{7}{*}{Karat3	}	&
1536		&	904	&	440	&	\textbf{302}\\
\cline{2-5}
	&	3072	&	2486	&	1272	&	\textbf{871}\\
\cline{2-5}
	&	6144	&	6104	&	3974		&	\textbf{2655}\\
\cline{2-5}
	&	12288	&	18463	&	11998	&	\textbf{8118}\\
\cline{2-5}
	&	24576	&	50311	&	37024	&	\textbf{24488}\\
\cline{2-5}
	&	49152	&	150469	&	117707	&	\textbf{76147}\\
\cline{2-5}
	&	98304	&	427281	&	357577	&	\textbf{231686}\\
\hline
\hline
\multirow{6}{*}{Karat5	}	&
2560	&	1770	&	1135	&	\textbf{724}	\\
\cline{2-5}
	&	5120	&	4640	&	3378	&	\textbf{2251}\\
\cline{2-5}
	&	10240	&	13814	&	10177		&	\textbf{6741}\\
\cline{2-5}
	&	20480	&	40911	&	30545	&	\textbf{20486}\\
\cline{2-5}
	&	40960	&	118334	&	97809	&	\textbf{63524}\\
\cline{2-5}
	&	81920	&	323997	&	295147	&	\textbf{192378}\\
\hline
\end{tabular}
\caption{Performance comparison for Algorithm~\ref{algo:karat3} and~\ref{algo:karat5}}\label{tab:perfKarat-3-5}
\end{center}
\end{table}

\begin{table*}[!h]
\begin{center}
\small
\begin{tabular}{cc}

	\begin{tabular}{|c|c|c|c|c|}
	\hline
		&	
	\multirow{2}{*}{size}	&	\multicolumn{3}{c|}{\# clock cycles}\\
	\cline{3-5}
		&&	
		\texttt{gf2x}&
			\texttt{AVX2}		&
				\texttt{AVX512}\\
	\hline
	\hline
	\multirow{6}{*}{Karat3(Karat3)	}	&
	4608	&	4129	&	2707	&	\textbf{1920}	\\
	\cline{2-5}
		&	9216	&	11261	&	8102	&	\textbf{5425}\\
	\cline{2-5}
		&	18432	&	34205	&	24498	&	\textbf{16340}\\
	\cline{2-5}
		&	36864	&	97586	&	76912	&	\textbf{51097}\\
	\cline{2-5}
		&	73728	&	270900	&	232362	&	\textbf{154065}\\
	\hline
	\hline
	\multirow{4}{*}{Karat5(Karat5)	}	&
	12800	&	19495	&	17988	&	\textbf{11471}	\\
	\cline{2-5}
		&	25600	&	56258	&	53132	&	\textbf{36417}\\
	\cline{2-5}
		&	51200	&	161740	&	159148		&	\textbf{107020}\\
	\cline{2-5}
		&	102400	&	438283	&	479256	&	\textbf{321361}\\
	\hline
	\end{tabular}

&

	\begin{tabular}{|c|c|c|c|c|}
	\hline
		&	
	\multirow{2}{*}{size}	&	\multicolumn{3}{c|}{\# clock cycles}\\
	\cline{3-5}
		&&	
		\texttt{gf2x}&
			\texttt{AVX2}		&
				\texttt{AVX512}\\
	\hline
	\hline
	\multirow{5}{*}{Karat3(Karat5)	}	&
	7680	&	8317	&	7075	&	\textbf{4496}	\\
	\cline{2-5}
		&	15360	&	25110	&	20670	&	\textbf{13845}\\
	\cline{2-5}
		&	30720	&	73222	&	63386		&	\textbf{42454}\\
	\cline{2-5}
		&	61440	&	209872	&	192557	&	\textbf{127336}\\
	\cline{2-5}
		&	122880	&	618612	&	597357	&	\textbf{388414}\\
	\hline
	\hline
	\multirow{5}{*}{Karat5(Karat3)	}	&
	7680	&	8242	&	7031	&	\textbf{4943}	\\
	\cline{2-5}
		&	15360	&	26364	&	20586	&	\textbf{13775}\\
	\cline{2-5}
		&	30720	&	72897	&	63653		&	\textbf{42713}\\
	\cline{2-5}
		&	61440	&	216280	&	190317	&	\textbf{129419}\\
	\cline{2-5}
		&	122880	&	619802	&	580575	&	\textbf{387763}\\
	\hline
	\end{tabular}
\\

\end{tabular}
\caption{Performance comparison for recursive Algorithms \ref{algo:karat3} and \ref{algo:karat5}}\label{tab:perfKaratCombined}
\end{center}
\end{table*}


\subsection{Conclusion}

In this section, we presented our \texttt{AVX512} implementation of recursive Karatsuba multiplication over $\mathbb F_2[X]$ for polynomial of degree at most 131071.



Our implementations show that the best \texttt{AVX512} approach is a 512 bit kernel, using a schoolbook algorithm at the 128 and 256 bit level, and Karatsuba at the highest level, that is for the 512 bit size operands.

Used in \texttt{AVX512} recursive Karatsuba multiplications of greater sizes, and in comparison with the \texttt{AVX2} software implementation of \cite{hqc20}, our implementations achieve a retired instruction number divided by roughly three, and speedups (in terms of clock cycle number) are from 29.5 \% to up to 39.1 \% (\texttt{KaratRec}, size 131072 bit), due to the lower IPC of the \texttt{AVX512} instruction set.

The same achievements have been reached in all the Karatsuba variants (split in 3, 5 parts, and combinations).



\section{Toom-Cook multiplication over $\mathbb F_2[X]$}
\label{sec:toom-cook}

In this section, we present the implementation issues of the Toom-Cook 3-5 approach applied to multiplication over $\mathbb F_2[X]$, especially with the \texttt{AVX512} instruction set. We refer the reader to Appendix \ref{app:TCGAlgo} page \pageref{app:TCGAlgo} for a general presentation of the Toom-Cook approach.

\subsection{Toom-Cook multiplication complexity}

There are several way to split the operands in the Toom-Cook approach:
\begin{itemize}
    \item The \texttt{Toom-Cook 3-5}, which splits the operands in 3 parts, and involves 5 elementary multiplications;
    \item The \texttt{Toom-Cook 4-7}, which splits the operands in 4 parts, and involves 7 elementary multiplications;
    \item The \texttt{Toom-Cook 5-9}, which splits the operands in 5 parts, and involves 9 elementary multiplications;
\end{itemize}
We present Table \ref{tab:compToom} the general results versus the polynomial degrees, and the corresponding operand size in 64 bit words.

As $\log_2(3) > \log_3(5) > \log_4(7) > \log_5(9)$, this implies that the multiplication number is decreasing while  increasing the split.
However, the hidden constant in the $\mathcal O$ notation is increasing, due to the more complex interpolation phase. This leads to a ``gray zone'' in which the different algorithms are very close to each other and this gray zone resides in our range for code based cryptographic protocols. In the \texttt{gf2x} implementation, a specific selection of the algorithm is made depending on each of the operand size (which can be different in this general purpose library). 

\begin{table*}[!h]
\begin{center}
\begin{tabular}{|c|c|c|c|}
\hline
mult.	&	Complexity	&	\multicolumn{2}{c|}{size range (\texttt{gf2x} \cite{gf2x})} \\
        &               &   operand degrees     &   \# 64 bit words  \\
\hline
\texttt{KaratRec}	&	$\mathcal O(t^{\log_2(3)})$	&	< 1343	& < 21\\
\hline
\texttt{Toom-Cook 3-5}	&	$\mathcal O(t^{\log_3(5)})$	&	1343 < degrees < 22143	&   21< $w$ <346 \\
\hline
\texttt{Toom-Cook 4-7}	&	$\mathcal O(t^{\log_4(7)})$	&	> 22143 & > 346	\\
\hline
\texttt{Toom-Cook 5-9}	&	$\mathcal O(t^{\log_5(9)})$	&	above	& above\\
\hline
\end{tabular}
\caption{Complexity of the Toom-Cook's multiplication variants}\label{tab:compToom}
\end{center}
\end{table*}

\subsection{Toom-Cook multiplication, implementation issues}
\label{sssec:impliss}

Let us recall that in order to multiply two binary polynomials $A$ and $B$ of degree at most $N-1$, we consider them as polynomials of degree at most $64t-1$ where $t=3n$ and $n$ ensures $t \geqslant \lceil N/64\rceil$. We now present how to choose $n$.

In the evaluation phase, the elementary products do not have the same operand size: $C(0), C(1)$ and $C(\infty)$ have operands of degree at most $64n-1$, while $C(x)$ and $C(x+1)$ have operands of degree at most $64n+2\cdot w-1$ (see Appendix \ref{app:TCGAlgo} page \pageref{app:TCGAlgo}). To take into account this characteristic, if the size of the elementary product is known, one has to set operands of size $n+2w/64$ 64-bit words, padding with zeros in order to use the same elementary product. Now, we can specify the value of $n$ mentioned Appendix \ref{app:TCGAlgo}: $n$ must be the minimum value ensuring $3n \geqslant \lceil N/64\rceil$ such that $n+2w/64$ is the size of the elementary multiplications computed during the evaluation phase. Thus, $n+2w/64$ is either a power of 2 in case of a Karatsuba multiplication (as seen section \ref{ssec:KaratRec}), or a value which complies with another Toom-Cook multiplication.

In the interpolation phase, since the divisions by $x$ and $(x+1)$ are exact, they can be implemented using the trick presented by Quercia and Zimmermann in \cite{Zimmermann08} and \cite{QuerciaZ03}. One takes advantage of the size of the polynomial to replace these divisions by a one word right shift for the division by $x$, and by a special multiplication by $(x+1)^{-1} \bmod X^d$, $d > $ degree of $C(x)$ and $d \equiv 0 \bmod w$.






In the second case, to evaluate $(x+1)^{-1} \bmod X^d = (X^w+1)^{-1} \bmod X^d$, notice that:

$$(X^w+1)^{-1} \bmod X^n = \sum^{d/w-1}_{i=0} X^{w\cdot i}\,.$$


\begin{exemple}
Here is a small toy example with polynomial over $\mathbb F_2[X]$ : let us divide by $X+1$ polynomial whose degree is less than $8$. One has $(X+1)^{-1} \mod X^8 = X^7+X^6+X^5+X^4+X^3+X^2+X+1$. Now, $\forall P(X)$ such as $(X+1)|P(X)$ and degree($P$)$<8$, $Q(X) = P(X)/(X+1)$ is computed as $P(X)\cdot(X+1)^{-1} \mod X^8$. Since this division is exact, the result is the exact quotient.

\noindent Let us set $P(X)=X^7 +X^5+X^4+ X$ :
$$\begin{array}{lcl}
	Q(X) &=& (X^7 +X^5+X^4+ X)\cdot (X+1)^{-1} \mod X^8\\
	&=& (X^7 +X^5+X^4+ X)\\&& \cdot (X^7+X^6+X^5+X^4+X^3+X^2+\\
	&& \hfill X+1) \bmod X^8\\
	&=&  X^6+X^5 +X^3+X^2+X.
\end{array}
$$
\normalsize

\end{exemple}

The vectorized implementation, while using 256-bit instructions, uses both values $w=64$ or $w=256$. We can even use $w=512$ in the case of \texttt{AVX512} platforms. Notice that the division by $x+1$ is cheaper in case of $w=256$ or $512$, while $n$ is slightly greater with $w=64$.


\medskip

To illustrate some Toom-Cook use cases, let us present some examples.

\begin{exemple}

In order to use elementary Karatsuba multiplications, and with $w=64$, let us consider $n$ such that $n+2w/64 = 2^8$. Thus, one has $64\cdot n+2\cdot w = 8192$, and this elementary multiplication proceeds polynomials of degree at most $8191$. We then have $n=254$, $t = 3\cdot n = 762$, thus building a Toom-Cook multiplication which can multiply polynomials of degree at most $762\cdot 64 -1 = 24191$.

\end{exemple}

\begin{exemple}

We now will use the previous multiplication in order to build a 1-recursive Toom-Cook multiplication, setting $w=256$. One now chooses $n$ such that $64n+2\cdot 256 = 24192$. This leads to $n = 370$ and $t = 1110$. This multiplication proceeds polynomials of degree at most $71039$:

\begin{itemize}
	\item the first split computes polynomials of degree at most $23679$;
	\item we build the operands for the evaluation phase, whose size is up to $24192$;
	\item we then call the Toom-Cook procedure of the previous example, which makes use of Karatsuba multiplications of size $8192$.
\end{itemize}
\end{exemple}

\begin{exemple}

We now  build a Toom-Cook multiplication, setting $w=512$ using a \texttt{Karat5} elementary multiplication of size $20480$. One now chooses $n$ such that $64n+2\cdot 512 = 20480$. This leads to $n = 304$ and $t = 912$. This multiplication proceeds polynomials of degree at most $58368$. This multiplication fits with the {hqc-256} protocol (see \cite{hqc20}), whose parameter $N=57669$, and for an \texttt{AVX512} implementation.

\end{exemple}

This can be adjusted for other sizes. To deal only with word shifts, this implies on the operand size that:

\begin{sloppypar}
\begin{itemize}
	\item elementary recursive Karatsuba multiplications are based on \texttt{PCLMULQDQ} or \texttt{VPCLMULQDQ} 64-bit elementary multiplications, whose size corresponds to the ones previously seen (KaratRec, Karat3, or Karat5...);
	\item the computation of $C(x)$ and $C(x+1)$ needs construction of operands whose size has to be the elementary multiplication size;
	\item the split has to take into account the size of the elementary multiplication operand, by diminishing the size of the split by two words. 
\end{itemize}

\end{sloppypar}

\subsubsection{ \texttt{Toom3Mult} implementations}

Three \texttt{Toom3Mult} versions have been implemented (see Table \ref{tab:perfToom3KR}):

\begin{itemize}
	\item with \texttt{KaratRec} elementary multiplications whose size is among 512, 1024, 2048, 4096, 8192, 16384 and 32768 bits;
	\item with \texttt{Karat3} elementary multiplications whose size is among 6144, 12288 and 32768 bits;
	\item with \texttt{Karat5} elementary multiplications whose size is among 5120, 10240, 20480, 40960 bits.
\end{itemize}

The platform and test procedure are the ones described Appendix \ref{app:perfProc} page \pageref{app:perfProc}.

\begin{table*}
\begin{center}
\begin{tabular}{|c|c|c|c|c|}
\hline
	&	
\multirow{2}{*}{size}	&	\multicolumn{3}{c|}{\# clock cycles}\\
\cline{3-5}
	&&	
	\texttt{gf2x}&
		\texttt{AVX2}		&
			\texttt{AVX512}\\
\hline
\hline
\multirow{3}{*}{Toom3Mult(KaratRec)	}	&
24192	&	47200	&	33501	&	\textbf{22238}\\
\cline{2-5}
	&	
48768	&	144960	&	104257	&	\textbf{69328}\\
\cline{2-5}
	&	
97920	&	427101	&	323719	&	\textbf{204596}\\
\hline
\hline
\multirow{3}{*}{Toom3Mult(Karat3)	}	&
18048	&	34019		&	22032	&	\textbf{14872}\\
\cline{2-5}
	&	
36480	&	94378	&	67451	&	\textbf{46940}\\
\cline{2-5}
	&	
73344	&	272891		&	206601	&	\textbf{137791}\\
\hline
\hline
\multirow{4}{*}{Toom3Mult(Karat5)	}	&
14976	&	25146		&	18570	&	\textbf{12664}\\
\cline{2-5}
	&	
30336	&	77548	&	57339	&	\textbf{38668}\\
\cline{2-5}
	&	
61056	&	211790	&	170316	&	\textbf{113783}	\\
\cline{2-5}
	&	
122496	&	622830	&	513595	&	\textbf{339387}\\
\hline
\end{tabular}
\caption{Performance comparison between \texttt{AVX2} and \texttt{AVX512} \texttt{Toom3} multiplications, with various elementary Karatsuba multiplications}\label{tab:perfToom3KR}
\end{center}
\end{table*}

The speedup of our \texttt{AVX512} implementation in comparison with the \texttt{AVX2} one is as follows:

\begin{sloppypar}
\begin{itemize}
	\item \texttt{Toom3} based on \texttt{KaratRec} (Toom3Mult(KaratRec)): the speedup starts from 33.5\% for the 48768 bit operand size up to 36.7\% for 97920 bit operand size;
	\item \texttt{Toom3} based on \texttt{Karat3} (Toom3Mult(Karat3)): the speedup starts from 30.4\% for 36480 bit operand size up to 33.3\% for the 73344 bit operand size;
	\item \texttt{Toom3} based on \texttt{Karat5} (Toom3Mult(Karat5)): the speedup starts from 32.4\% for the 14976 bit operand size up to 33.9\% for 122496 bit operand size.
\end{itemize}

\end{sloppypar}


\subsection{\texttt{Toom3} vs \texttt{Karat3} comparison}

The Toom-Cook multiplication presented above\break (\texttt{Toom3}) and the \texttt{Karat3} multiplication (see Algorithm \ref{algo:karat3} page \pageref{algo:karat3}) look similar, both splitting the operands in three parts. However, the \texttt{Toom3} needs 5 elementary multiplications while the \texttt{Karat3} requires 6. In the \texttt{gf2x} library \cite{gf2x}, the \texttt{Toom-3} multiplication is used above the threshold of 21 64-bit words, i.e. polynomial of degree 1343. 

In our \texttt{AVX2} and \texttt{AVX512} implementations, we compare the clock cycle numbers of multiplications using the same elementary Karatsuba multiplication. Consequently, the operand size in the \texttt{Toom3} case is slightly lower. In Table \ref{tab:perfCompToom3K3}, the clock cycle numbers of the \texttt{Toom3} multiplication are lower for the considered sizes, and the speedup starts from 8\% for the smallest (about 6000 bit operand size)  up to 11.7\% for the bigger sizes, the maximum potential speedup being theoretically 16.7\%. Indeed, the costliest interpolation and reconstruction phase of the \texttt{Toom3} approach lowers the speedup for the smallest sizes.

\begin{table*}
\begin{center}
\begin{tabular}{|c|c|c||c|c||c|c|}

\hline
	\multicolumn{3}{|c|}{}		&	\multicolumn{4}{c|}{ \# clock cycles}\\
	\multicolumn{3}{|c|}{multiplication size}	&	\multicolumn{2}{c||}{\texttt{AVX2}}	&	\multicolumn{2}{c|}{\texttt{AVX512}}\\
\hline
Elt. Karat. size	&	\texttt{Karat3}	&	\texttt{Toom3}	&	\texttt{Karat3}	&	\texttt{Toom3}	&	\texttt{Karat3}	&	\texttt{Toom3}\\
\hline
\hline
2048	&	6144	&	5760	&	4150	&	\bf 3987	&	2655	&	-	\\
\hline
3072	&	9216	&	8832	&	8405	&	\bf 7726	&	5425	&	-	\\
\hline
4096	&	12288	&	11998	&	12508	&	\bf 11480	&	8118	&	-	\\		
\hline
6144	&	18432	&	18048	&	25264	&	\bf 22032	&	16340	&	\bf 14872	\\
\hline
8192	&	24576	&	24192	&	37024	&	\bf 33501	&	24488	&	\bf 22238	\\
\hline
12288	&	36864	&	36480	&	79727	&	\bf 67451	&	51097	&	\bf 46940	\\
\hline
16384	&	49152	&	48768	&	117707	&	\bf 104257	&	76147	&	\bf 69328	\\
\hline
24576	&	73728	&	73344	&	241548	&	\bf 206601	&	154065	&	\bf 137791	\\
\hline
32768	&	98304	&	97920	&	357577	&	\bf 323719	&	231686	&	\bf 204596	\\
\hline
\end{tabular}
\caption{Performance comparison between \texttt{AVX2} and \texttt{AVX512}  \texttt{Toom3Mult} and \texttt{Karat3} multiplications}\label{tab:perfCompToom3K3}
\end{center}
\end{table*}


\subsection{Conclusion}

\begin{sloppypar}
In this section, we presented our \texttt{AVX512} implementation of Toom-Cook multiplication over $\mathbb F_2[X]$ for polynomial of degree at most 122496, based on various Karatsuba version elementary \texttt{AVX512} multiplications (\texttt{KaratRec}, \texttt{Karat3} and \texttt{Karat5}).
\end{sloppypar}

The speedup between \texttt{AVX2} and our \texttt{AVX512} implementation is again up to nearly 37\% (97920 bit size, \texttt{Toom3Mult(KaratRec)}), as it has already been observed with our \texttt{AVX512} Karatsuba multiplications. Thus, the same remark can be done about the potential in terms of future results.


\section{HQC multiplications}
\label{sec:HQC}

In this section, we present the application of our \texttt{AVX2} and \texttt{AVX512} Toom-Cook multiplications in the context of the HQC protocol.

The sizes of the HQC multiplications are (see \cite{hqc20}):

\begin{itemize}
	\item \texttt{hqc-128} : \texttt{PARAM\_N} $=17669$;
	\item \texttt{hqc-192} : \texttt{PARAM\_N} $=35851$;
	\item \texttt{hqc-256} : \texttt{PARAM\_N} $=57637$.
\end{itemize}

\begin{sloppypar}
In the NIST submission \cite[updated submission package (round 3), 2020/10/01]{hqc20}, the \texttt{hqc-128} and the \texttt{hqc-192} implementations make use of the \texttt{Karat3} multiplication based on elementary \texttt{Karat3} multiplications, while the \texttt{hqc-256} is a Toom-Cook multiplication (3 part operand split) based on a \texttt{Karat5} elementary multiplication. The reader may notice that the operand size is different in the \texttt{hqc-256} case, in comparison with the Toom3Mult(Karat5) version presented in Table \ref{tab:perfToom3KR} page \pageref{tab:perfToom3KR}, whose operand size is 61056 bits, while the \texttt{AVX2} operand size is 59904 bits, and the \texttt{AVX512} one is 58368 bits, see Table \ref{tab:NRperfs}. This is due to the word size considered in the Toom-Cook implementation: the first version of table \ref{tab:perfToom3KR} has a word size $w = 64$ bits, while the \texttt{AVX2} version has $w = 256$ and we chose $w = 512$ for the \texttt{AVX512} version. The new sizes fit the \texttt{hqc-256} \texttt{PARAM\_N} $=57637$, allowing a slightly better performance.
\end{sloppypar}

The platform and test procedure are the ones described Appendix \ref{app:perfProc} page \pageref{app:perfProc}. We kept the compiler flags used in the NIST submission \cite[updated submission package (round 3), 2020/10/01]{hqc20}:
\\
\texttt{-O3 -funroll-all-loops -flto -march=tigerlake} \\ for \texttt{AVX512} versions;
\\\texttt{-O3 -funroll-all-loops -flto -mavx -mavx2 -mbmi -mpclmul} for \texttt{AVX2} versions.
\\

In Table \ref{tab:NRperfs}, we sum-up the performances of the respective multiplications, from Tables \ref{tab:perfToom3KR} page \pageref{tab:perfToom3KR} and \ref{tab:perfCompToom3K3} page \pageref{tab:perfCompToom3K3}, along with the specific \texttt{hqc-256} multiplications results, in the context of HQC protocol.

\begin{table*}
\begin{center}
	\begin{tabular}{|c|c|c|c|}
		\hline
			\multicolumn{2}{|c|}{}	&	\multicolumn{2}{c|}{\bf This work}	\\
		\cline{2-4}
			&	NIST release \cite{hqc20}	&	improved \texttt{AVX2}	&	\texttt{AVX512}	\\
		\hline
		\hline
		\texttt{hqc-128}	&	Karat9 ($N=18432$)	&	\multicolumn{2}{c|}{Toom3Karat3 ($N=18048, w=64$)}	\\
		\cline{2-4}
		\texttt{PARAM\_N} $=17669$ &	24498	&	22032	&	\bf 14872	\\
		\hline
		\hline
		\texttt{hqc-192}	&	Karat9 ($N=36864$)	&	\multicolumn{2}{c|}{Toom3Karat3 ($N=36480, w=64$)}	\\
		\cline{2-4}
		\texttt{PARAM\_N} $=35851$ &	76912	&	67451	&	\bf 46940	\\
		\hline
		\hline
		\texttt{hqc-256}	&	\multicolumn{2}{c|}{Toom3Karat5 ($N=59904, w=256$)}	&	Toom3Karat5 ($N=58368, w=512$)	\\
		\cline{2-4}
		\texttt{PARAM\_N} $=57637$	&	\multicolumn{2}{c|}{168975}	&	\bf 110568	\\
\hline
	\end{tabular}
\caption{\texttt{AVX2} and \texttt{AVX512} multiplication version performances, clock cycles numbers, in the context of the HQC protocol  }\label{tab:NRperfs}
\end{center}
\end{table*}

\subsection{HQC implementation performances}

This leads to the performances and also the speedups brought by our work in comparison with the NIST release \cite[updated submission package (round 3),\break 2020/10/01]{hqc20}, provided in Table \ref{tab:HQCperfs}.

As a consequence of this work, the \texttt{AVX2} Toom3 multiplication based on Karat3 is now part of the last release of HQC (2021/06/06).

Table \ref{tab:HQCAVX512perfs} provides the HQC performances when using our \texttt{AVX512} multiplications in the 2021/06/06 release. One may notice that column one of this Table is different of the corresponding one of Table \ref{tab:HQCperfs} because other parts of the HQC source code has been updated between both releases.

A maximum 11.8\% speedup for the HQC implementation is reached with the \texttt{hqc-256} Keygen. 

\begin{table*}[!h]
\begin{center}
	\begin{tabular}{|c|c|c|c|}
		\hline
		\multicolumn{3}{|c|}{}&{\bf This work}	\\
		\cline{3-4}
			\multicolumn{2}{|c|}{\# clock cycles}&	NIST release	&	\multirow{2}{*}{improved \texttt{AVX2}}	\\
			\multicolumn{2}{|c|}{}&	2020/10/01	&		\\
		\hline
		\hline
		\multirow{3}{*}{\texttt{hqc-128}}	&	Keygen	&	111073	&	110009 (-1.0 \%)		\\
		\cline{2-4}
			&	Encaps	&	185741	&	181154 (-2.4 \%)		\\
		\cline{2-4}
			&	Decaps	&	344154	&	337594 (-2.0 \%)		\\
		\hline
		\hline
		\multirow{3}{*}{\texttt{hqc-192}}	&	Keygen	&	250184	&	239908 (-5.3 \%)	\\
		\cline{2-4}
			&	Encaps	&	430689	&	410090 (-4.8 \%)		\\
		\cline{2-4}
			&	Decaps	&	722899	&	696779 (-3.6 \%)	\\
		\hline
	\end{tabular}
\caption{HQC performances, \texttt{AVX2} clock cycles  numbers }\label{tab:HQCperfs}
\end{center}
\end{table*}

\begin{table*}[!h]
\begin{center}
	\begin{tabular}{|c|c|c|c|}
		\hline
		\multicolumn{3}{|c|}{}&{\bf This work}	\\
		\cline{3-4}
			\multicolumn{2}{|c|}{\# clock cycles}&	NIST release	&	\multirow{2}{*}{\texttt{AVX512}}	\\
			\multicolumn{2}{|c|}{}&	2021/06/06	&		\\
		\hline
		\hline
		\multirow{3}{*}{\texttt{hqc-128}}	&	Keygen	&	70171	&	64825 (-7.6 \%)		\\
		\cline{2-4}
			&	Encaps	&	1723219	& 158377 (-8.1 \%)		\\
		\cline{2-4}
			&	Decaps	&	311434	&	300661 (-3.5 \%)		\\
		\hline
		\hline
		\multirow{3}{*}{\texttt{hqc-192}}	&	Keygen	&	168397	&	154486 (-8.3 \%)	\\
		\cline{2-4}
			&	Encaps	&	395367	&	361838 (-8.5 \%)		\\
		\cline{2-4}
			&	Decaps	&	646313	&	617853 (-4.4 \%)	\\
		\hline
		\hline
		\multirow{3}{*}{\texttt{hqc-256}}	&	Keygen	&	338137	&	298331 (-11.8 \%)	\\
		\cline{2-4}
			&	Encaps	&	768537	&	680602 (-11.4 \%)		\\
		\cline{2-4}
			&	Decaps	&	1290132	&	1194526 (-7.4 \%)	\\
		\hline
	\end{tabular}
\caption{HQC performances, \texttt{AVX512} clock cycles  numbers }\label{tab:HQCAVX512perfs}
\end{center}
\end{table*}





\section{Conclusion}
\label{sec:conclusion}
In this paper, we considered the software \texttt{AVX512} implementation of polynomial multiplication over $\mathbb F_2[X]$, using the vectorized 64-bit polynomial multiplication instruction \texttt{VPCLMULQDQ}. We studied the different combinations of schoolbook/Karatsuba constructions for the kernels up to 512 bit operands. We then implemented two different approaches: one based on the Karatsuba subquadratic approach and the other on the Toom-Cook approach. These implementations are competitive in comparison with state-of-the art general purpose library, HQC submissions, and other \texttt{AVX512} software implementations of \cite{NDruckerGK18,DruckerGK20}. While the retired instruction count is divided by roughly three compared to the corresponding \texttt{AVX2} implementations, we achieved a speedup up to nearly 40\%, in terms of clock cycle numbers.

We implemented our approaches in the HQC protocol by patching the NIST submission released in october 2020, in order to experiment the potential benefits, and this leads to speedups up to 11.8\% (\texttt{hqc-256} Keygen.).

All the implementations of this work are available on github\footnote{\url{https://github.com/arithcrypto/AVX512PolynomialMultiplication}}.

\textbf{Funding} This work has been partially funded by TPM Metropol  (AAP2020-IPOCRAS project).


\bibliographystyle{plainurl}
\bibliography{biblio.bib}

\begin{appendix}
\normalsize

\section{Karatsuba algorithms}
\label{app:Karat}

We reproduce here the Karatsuba algorithms:

\begin{itemize}
    \item Algorithm \ref{algo:karatrec} reproduces the recursive multiplication with a two halves split, from \cite{NegreR13};
    \item Algorithm \ref{algo:karat3} shows the three parts split corresponding approach;
    \item Algorithm \ref{algo:karat5} shows the five parts split corresponding approach;
\end{itemize}

\begin{algorithm}
\caption{\texttt{KaratRec}(A,B,t), from \cite{NegreR13}} \label{algo:karatrec}
\begin{center}
\begin{algorithmic}[1] 
\REQUIRE $A$ and $B$ on $t=2^r$ computer words. 
\ENSURE  $R=A \times B$
\IF{ $t = 1$ }
\RETURN(  $Mult64(A,B)$ )
\ELSE
\STATE // Split  in two halves of word size $t/2$.  
\STATE $A=A_0+x^{64t/2}A_1$ 
\STATE $ B=B_0+x^{64t/2}B_1$
\STATE // Recursive multiplications
\STATE $R_0 \gets \texttt{KaratRec}(A_0,B_0,t/2)$
\STATE $R_1 \gets \texttt{KaratRec}(A_1,B_1,t/2)$
\STATE $R_2 \gets \texttt{KaratRec}(A_0+A_1,B_0+B_1,t/2)$
\STATE // Reconstruction
\STATE $R \gets R_0 + (R_0+R_1+R_2)X^{64t/2}+R_1X^{64t}$
\RETURN($R$)
\ENDIF 
\end{algorithmic}
\end{center}
\end{algorithm}

\begin{algorithm}[!h]
\caption{\texttt{Karat3}(A,B,t), from \cite{AWeimerskirchP06}} \label{algo:karat3}
\begin{center}
\begin{algorithmic}[1] 
\REQUIRE $A$ and $B$ on $t=3\times 2^r$ computer words. 
\ENSURE  $R=A \times B$
\STATE // Split  in three thirds of word size $t/3$.  
\STATE $A=A_0+x^{64t/3}A_1 +x^{2\times 64t/3}A_2$
\STATE $ B=B_0+x^{64t/3}B_1+x^{2\times 64t/3}B_2$
\STATE // Recursive multiplications
\STATE $R_0 \gets \texttt{KaratRec}(A_0,B_0,t/3)$
\STATE $R_1 \gets \texttt{KaratRec}(A_1,B_1,t/3)$
\STATE $R_2 \gets \texttt{KaratRec}(A_2,B_2,t/3)$
\STATE $R_3 \gets \texttt{KaratRec}(A_0+A_1,B_0+B_1,t/3)$
\STATE $R_4 \gets \texttt{KaratRec}(A_0+A_2,B_0+B_2,t/3)$
\STATE $R_5 \gets \texttt{KaratRec}(A_1+A_2,B_1+B_2,t/3)$
\STATE // Reconstruction
\STATE $R \gets R_0 + (R_0+R_1+R_3)X^{64t/3} + (R_0+R_1+R_2+R_4)X^{2\times 64t/3}  + (R_1+R_2+R_5)X^{64t}   + R_2X^{4\times 64t/3} $
\RETURN($R$)
\end{algorithmic}
\end{center}

\end{algorithm}
\begin{algorithm}[!h]
\caption{\texttt{Karat5}(A,B,t), from \cite{AWeimerskirchP06}} \label{algo:karat5}
\begin{center}
\begin{algorithmic}[1] 
\REQUIRE $A$ and $B$ on $t=5\times 2^r$ computer words. 
\ENSURE  $R=A \times B$
\STATE // Split  in five parts of word size $t/5$.  
\STATE $A=A_0+x^{64t/5}A_1 +x^{2\times 64t/5}A_2 +x^{3\times 64t/5}A_3+x^{4\times 64t/5}A_4$
\STATE $ B=B_0+x^{64t/5}B_1+x^{2\times 64t/5}B_2 +x^{3\times 64t/5}B_3+x^{4\times 64t/5}B_4 $
\STATE // Recursive multiplications
\STATE $R_0 \gets \texttt{KaratRec}(A_0,B_0,t/5)$
\STATE $R_1 \gets \texttt{KaratRec}(A_1,B_1,t/5)$
\STATE $R_2 \gets \texttt{KaratRec}(A_2,B_2,t/5)$
\STATE $R_3 \gets \texttt{KaratRec}(A_3,B_3,t/5)$
\STATE $R_4 \gets \texttt{KaratRec}(A_4,B_4,t/5)$

\STATE $R_{01} \gets \texttt{KaratRec}(A_0+A_1,B_0+B_1,t/5)$
\STATE $R_{02} \gets \texttt{KaratRec}(A_0+A_2,B_0+B_2,t/5)$
\STATE $R_{03} \gets \texttt{KaratRec}(A_0+A_3,B_0+B_3,t/5)$
\STATE $R_{04} \gets \texttt{KaratRec}(A_0+A_4,B_0+B_4,t/5)$

\STATE $R_{12} \gets \texttt{KaratRec}(A_1+A_2,B_1+B_2,t/5)$
\STATE $R_{13} \gets \texttt{KaratRec}(A_1+A_3,B_1+B_3,t/5)$
\STATE $R_{14} \gets \texttt{KaratRec}(A_1+A_4,B_1+B_4,t/5)$

\STATE $R_{23} \gets \texttt{KaratRec}(A_2+A_3,B_2+B_3,t/5)$
\STATE $R_{24} \gets \texttt{KaratRec}(A_2+A_4,B_2+B_4,t/5)$

\STATE $R_{34} \gets \texttt{KaratRec}(A_3+A_4,B_3+B_4,t/5)$

\STATE // Reconstruction
\STATE $R \gets R_0 + (R_0+R_1+R_{01})X^{64t/5} + (R_0+R_1+R_2+R_{02})X^{2\times 64t/5}  + (R_0+R_1+R_2+R_3+R_{03}+R_{12})X^{3\times 64t/5}   + (R_0+R_1+R_2+R_3+R_4+R_{04}+R_{13})X^{4\times 64t/5} + (R_1+R_2+R_3+R_4+R_{14}+R_{23})X^{64t} + (R_3+R_2+R_4+R_{24})X^{6\times 64t/5}  + (R_3+R_4+R_{34})X^{7\times 64t/5} + R_4X^{8\times 64t/5}  $
\RETURN($R$)
\end{algorithmic}
\end{center}
\end{algorithm}

\lstset{language=C,backgroundcolor=\color{lightgray!30}}

\section{Source code for 256-bit operand size multiplication}
\label{app:256bitMult}

\subsection{Source code for the $4\times 4$ 256 bit multiplication of this work based on the schoolbook approach}

We present here our schoolbook \texttt{AVX512} implementation of the 256 bit operand size multiplication, with comments and explanations.

\scriptsize
\begin{lstlisting}
__m512i mask_middle= (__m512i){0x0UL,
            0xffffffffffffffffUL,
	    0xffffffffffffffffUL,0x0UL,0x0UL,
	    0xffffffffffffffffUL,
	    0xffffffffffffffffUL,0x0UL};

__m512i idx_b=(__m512i){0x0UL,0x1UL,0x2UL,0x3UL
                ,0x2UL,0x3UL,0x0UL,0x1UL};
__m512i idx_1=(__m512i){0x0UL,0x1UL,0x8UL,0x9UL,
                0x2UL,0x3UL,0xaUL,0xbUL};
__m512i idx_2=(__m512i){0x0UL,0x1UL,0x6UL,0x7UL,
                0x2UL,0x3UL,0x4UL,0x5UL};
__m512i idx_3=(__m512i){0x0UL,0x1UL,0x4UL,0x5UL,
                0x2UL,0x3UL,0x6UL,0x7UL};
__m512i idx_4=(__m512i){0x8UL,0x0UL,0x1UL,0x2UL,
                0x3UL,0x4UL,0x5UL,0x8UL};
__m512i idx_5=(__m512i){0x8UL,0x8UL,0x8UL,0x6UL,
                0x7UL,0x8UL,0x8UL,0x8UL};
__m512i idx_6=(__m512i){0x0UL,0x0UL,0x4UL,0x5UL,
                0xcUL,0xdUL,0x0UL,0x0UL};
__m512i idx_7=(__m512i){0x0UL,0x0UL,0x6UL,0x7UL,
                0xeUL,0xfUL,0x0UL,0x0UL};

\end{lstlisting}
\normalsize

\bigskip

\begin{sloppypar}
These lines define the constant indexes for the \texttt{\_mm512\_permutexvar\_epi64} and  \texttt{\_mm512\_permutex2var\_epi64} instructions. These instructions are explained on the fly.
\end{sloppypar}

\bigskip

\scriptsize
\begin{lstlisting}
void mult_256_256_512(__m512i * Out,
            const __m256i * A256,
            onst __m256i * B256)
{

 __m512i A512, B512 ;
 __m512i R0_512,R1_512,R2_512, R3_512,
         middle, tmp;

 A512 =_mm512_broadcast_i64x4(*A256);
 tmp  =_mm512_broadcast_i64x4(*B256);
 B512 =_mm512_permutexvar_epi64 (idx_b, tmp);
	
\end{lstlisting}
	\normalsize

\bigskip

\begin{sloppypar}
The \texttt{\_mm512\_broadcast\_i64x4(*A256)} instruction duplicates the 256 bits of \texttt{*A256~} in the \texttt{A512~} register, the same for the  \texttt{*B256}.
\end{sloppypar}

\begin{sloppypar}
The \texttt{\_mm512\_permutexvar\_epi64 (idx\_b, tmp)} spreads the 64 bit words following the index \texttt{idx\_b}. This allows to shuffle the 64 bit words of \texttt{*B256~} in the \texttt{B512~} register, in order to prepare the elementary multiplications.
\end{sloppypar}

We thus have:

\texttt{A512~}$ \leftarrow \{a_3,a_2,a_1,a_0,a_3,a_2,a_1,a_0\}$

\texttt{tmp~~}$ \leftarrow \{b_3,b_2,b_1,b_0,b_3,b_2,b_1,b_0\}$

\texttt{B512~}$ \leftarrow \{b_1,b_0,b_3,b_2,b_3,b_2,b_1,b_0\}$

\bigskip

\scriptsize
\begin{lstlisting}
 R0_512=_mm512_clmulepi64_epi128(A512,B512,0x00);
 R1_512=_mm512_clmulepi64_epi128(A512,B512,0x10);
 R2_512=_mm512_clmulepi64_epi128(A512,B512,0x01);
 R3_512=_mm512_clmulepi64_epi128(A512,B512,0x11);
\end{lstlisting}
\normalsize

We now compute all the elementary 64 bit multiplications, providing all the 128 bit results as follows:
	
	\texttt{R0~}$ \leftarrow \{a_{2}\times b_{0},a_{0}\times b_{2},a_{2}\times b_{2},a_{0}\times b_{0}\}$
	
	\texttt{R1~}$ \leftarrow \{a_{2}\times b_{1},a_{0}\times b_{3},a_{2}\times b_{3},a_{0}\times b_{1}\}$
	
	\texttt{R2~}$ \leftarrow \{a_{3}\times b_{0},a_{1}\times b_{2},a_{3}\times b_{2},a_{1}\times b_{0}\}$
	
	\texttt{R3~}$ \leftarrow \{\underbrace{a_{3}\times b_{1}}_{128 bits},\underbrace{a_{1}\times b_{3}}_{128 bits},\underbrace{a_{3}\times b_{3}}_{128 bits},\underbrace{a_{1}\times b_{1}}_{128 bits}\}$

\bigskip

\scriptsize
\begin{lstlisting}
 tmp = _mm512_permutex2var_epi64
        (R0_512,idx_1,R3_512);
\end{lstlisting}
\normalsize

\bigskip

The \texttt{tmp~} register now contains all the $a_i\times b_i$ elementary products coming form the \texttt{R0\_512} and \texttt{R3\_512} registers:

	\texttt{tmp~}$ \leftarrow \{a_{3}\times b_{3},a_{2}\times b_{2},a_{1}\times b_{1},a_{0}\times b_{0}\}$

It remains now to compute the middle part of the result to be added to \texttt{tmp}, in order to get the final result.

\scriptsize
\begin{lstlisting}
 middle = _mm512_permutexvar_epi64(idx_2, R1_512);
 middle ^=_mm512_permutexvar_epi64(idx_3, R2_512);
\end{lstlisting}
\normalsize

\begin{sloppypar}
The \texttt{middle~} register now contains the addition (XOR) between \texttt{R1\_512} and \texttt{R2\_512}, reordered with the \texttt{\_mm512\_permutexvar\_epi64}:
\end{sloppypar}

	\texttt{middle~}$ \leftarrow \{a_{0} b_{3}\oplus a_{3} b_{0},a_{2} b_{3}\oplus a_{3} b_{2},$
	
	\hspace{1cm}$~~~~~a_{1} b_{2}\oplus a_{2} b_{1},a_{1} b_{0}\oplus a_{0} b_{1}\}$

\bigskip

\scriptsize
\begin{lstlisting}
 tmp ^= _mm512_permutex2var_epi64
        (middle,idx_4,idx_b);
 tmp ^= _mm512_permutex2var_epi64
        (middle,idx_5,idx_b);
\end{lstlisting}
\normalsize

\bigskip

The \texttt{tmp~} register is added (XOR) with the elementary products of the \texttt{middle~} register, and nearly contains the result, except some of the products of the middle part:

\noindent\scriptsize\begin{tabular}{cc}
	
	\texttt{tmp~}$\gets$	&
	
		\begin{tabular}{cccccccc}
			\multicolumn{2}{c|}{$a_{3}\times b_{3}$~~~}	&	\multicolumn{2}{c|}{~~~$a_{2}\times b_{2}$~~~~~}		&\multicolumn{2}{c|}{~~$a_{1}\times b_{1}$~~~}	&	\multicolumn{2}{c}{~~$a_{0}\times b_{0}$}	\\
	
			$\oplus$	&	\multicolumn{2}{c|}{$a_{2} b_{3}\oplus a_{3} b_{2}$}	&	\multicolumn{2}{c|}{$a_{1} b_{2}\oplus a_{2} b_{1}$}	&	\multicolumn{2}{c}{$a_{1} b_{0}\oplus a_{0} b_{1}$}	&\\

			$\oplus$	&&&	\multicolumn{2}{c}{$a_{0} b_{3}\oplus a_{3} b_{0}$}		&&&\\
		\end{tabular}
			\\
	\end{tabular}

\bigskip

\scriptsize
\begin{lstlisting}
 middle = _mm512_permutex2var_epi64(R0_512,
                              idx_6,R3_512); 
 middle ^= _mm512_permutex2var_epi64(R0_512,
                               idx_7,R3_512);
\end{lstlisting}
\normalsize

\bigskip

The remaining products of the middle part to be added with \texttt{tmp} are put in place in the \texttt{middle~} register:

	\texttt{middle~}$ \leftarrow  \{\underbrace{0x0UL}_{128 bits},a_{1} b_{3}\oplus a_{3} b_{1},a_{0} b_{2}\oplus a_{2} b_{0},\underbrace{0x0UL}_{128 bits}\}$

\scriptsize
\begin{lstlisting}

	*Out = tmp^middle;
}
	
\end{lstlisting}
\normalsize

\bigskip

	\texttt{Out} gets the final reconstruction :

\medskip
\scriptsize
\noindent\begin{tabular}{ccc}

    \texttt{Out}	&
    $\gets$	&
    
    \begin{tabular}{cccccccc}
    	\multicolumn{2}{c|}{$a_{3}\times b_{3}$}	&	\multicolumn{2}{c|}{$a_{2}\times b_{2}$}	&	\multicolumn{2}{c|}{$a_{1}\times b_{1}$}	&	\multicolumn{2}{c}{$a_{0}\times b_{0}$}	\\
    
    	$\oplus$	&	\multicolumn{2}{c|}{$a_{2} b_{3}\oplus a_{3} b_{2}$}	&	\multicolumn{2}{c|}{$a_{1} b_{2}\oplus a_{2} b_{1}$}	&	\multicolumn{2}{c}{$a_{1} b_{0}\oplus a_{0} b_{1}$}	&\\
    
    	$\oplus$	&&	\multicolumn{2}{c|}{$a_{1} b_{3}\oplus a_{3} b_{1}$}	&	\multicolumn{2}{c}{$a_{0} b_{2}\oplus a_{2} b_{0}$}	&&\\
    	
    	$\oplus$	&&&	\multicolumn{2}{l}{\hglue -3mm$a_{0} b_{3}\oplus a_{3} b_{0}$}		&&&\\
    \end{tabular}
	\\
\end{tabular}
\normalsize


\section{Source code for 512-bit operand size multiplications}
\label{app:512bitMult}


\begin{sloppypar}
We detail now the source code of the \texttt{karat\_mult\_1\_512\_SB} procedure.

First, the preamble declares the constant indexes for the \texttt{\_mm512\_permutex2var\_epi64} and \texttt{\_mm512\_permutexvar\_epi64} instructions.
\end{sloppypar}

\scriptsize
\begin{lstlisting}

inline static void karat_mult_1_512(__m512i * C,
         const __m512i * A, const __m512i * B)
{

 const __m512i perm_al = (__m512i){0x0UL,0x1UL,
                            0x0UL,0x1UL,
                            0x2UL,0x3UL,
                            0x2UL,0x3UL};
 const __m512i perm_ah = (__m512i){0x4UL,0x5UL,
                            0x4UL,0x5UL,
                            0x6UL,0x7UL,
                            0x6UL,0x7UL};
 const __m512i perm_bl = (__m512i){0x0UL,0x1UL,
                            0x2UL,0x3UL,
                            0x0UL,0x1UL,
                            0x2UL,0x3UL};
 const __m512i perm_bh = (__m512i){0x4UL,0x5UL,
                            0x6UL,0x7UL,
                            0x4UL,0x5UL,
                            0x6UL,0x7UL};
 const __m512i mask_R1 = _mm512_set_epi64
                          (6,7,4,5,2,3,0,1);
 const __m512i perm_h = (__m512i){0x4UL,0x5UL,
                            0x0UL,0x1UL,
                            0x2UL,0x3UL,
                            0x6UL,0x7UL};
 const __m512i perm_l = (__m512i){0x0UL,0x1UL,
                            0x4UL,0x5UL,
                            0x6UL,0x7UL,
                            0x2UL,0x3UL};
 const __m512i mask = _mm512_set_epi64
                         (15,14,13,12,3,2,1,0);

\end{lstlisting}
\normalsize

Next, we compute the registers \texttt{al}, \texttt{ah}, \texttt{bl}, \texttt{bh}, \texttt{sa} and \texttt{sb} so that they contain the split parts for the 256 bit operands, and the corresponding sums for the Karatsuba 256 bit middle multiplication.

\scriptsize
\begin{lstlisting}

 __m512i al = _mm512_permutexvar_epi64
                        (perm_al, *A );
 __m512i ah = _mm512_permutexvar_epi64
                        (perm_ah, *A );
 __m512i bl = _mm512_permutexvar_epi64
                        (perm_bl, *B );
 __m512i bh = _mm512_permutexvar_epi64
                        (perm_bh, *B );
    
 __m512i sa = al^ah;
 __m512i sb = bl^bh;
	
\end{lstlisting}
\normalsize

We compute now the three 256 bit multiplications in order to prepare the 512 bit registers \texttt{cl}, \texttt{ch} and \texttt{cm} containing their results.

\scriptsize
\begin{lstlisting}

 // first multiplication 256 : AlBl

 __m512i R0_512=_mm512_clmulepi64_epi128
                        (al,bl,0x00);
 __m512i R1_512=_mm512_clmulepi64_epi128
                        (al,bl,0x01);
 __m512i R2_512=_mm512_clmulepi64_epi128
                        (al,bl,0x10);
 __m512i R3_512=_mm512_clmulepi64_epi128
                        (al,bl,0x11);

 R1_512 = _mm512_permutexvar_epi64
                ( mask_R1 , R1_512^R2_512 ) ;

 __m512i l =  _mm512_mask_xor_epi64( R0_512 ,
                        0xaa , R0_512 , R1_512 ) ;
 __m512i h =  _mm512_mask_xor_epi64( R3_512 ,
                        0x55 , R3_512 , R1_512 ) ;

\end{lstlisting}
\normalsize

These lines computes four 128 bit operand size multiplications in parallel, using a schoolbook approach. This procedure acts like the \texttt{mul128x4} procedure of Drucker \emph{et al.} \cite{DruckerGK20}, which is based on the Karatsuba algorithm\footnote{We do not present in detail our variant \texttt{karat\_mult\_1\_512} based on the \texttt{mul128x4}, however, we refer the reader to their paper \cite{DruckerGK20} for its presentation.}.

The 512 bit registers \texttt{l} and \texttt{h} now contains the four elementary 256 bit results.

\scriptsize
\begin{lstlisting}

 __m512i cl = _mm512_permutex2var_epi64(l,mask,h);
 l = _mm512_permutexvar_epi64(perm_l, l );	
 h = _mm512_permutexvar_epi64(perm_h, h );

 __m512i middle = _mm512_maskz_xor_epi64(0x3c,h,l);

 cl ^= middle;
	
\end{lstlisting}
\normalsize

This is the schoolbook reconstruction for the first 256 bit multiplication. The register \texttt{cl} now contains the 512 bit result of \texttt{al$\times$bl}.

We now compute the same the two remaining 256 bit operand size multiplications:

\scriptsize
\begin{lstlisting}
	
 // second multiplication 256 : AhBh
...
 ch ^= middle;


 // third multiplication 256 : SASB
...
 cm ^= middle^cl^ch;
 
\end{lstlisting}
\normalsize

The register \texttt{ch} now contains the 512 bit result of \texttt{ah$\times$bh}.

The result \texttt{cm} is directly added (XOR) to the other results \texttt{cl} and \texttt{ch} in order to prepare the final Karatsuba reconstruction, and \texttt{cm} now contains the 512 bit result of \texttt{(sa$\times$sa)$\oplus$cl$\oplus$ch}:

\scriptsize
\begin{lstlisting}
	
 // final reconstruction (Karatsuba)
	
 const __m512i perm_cm = (__m512i){0x4UL,0x5UL,
            0x6UL,0x7UL,0x0UL,0x1UL,0x2UL,0x3UL};
            
 cm = _mm512_permutexvar_epi64(perm_cm, cm );	
	
 C[0]= _mm512_mask_xor_epi64(cl,0xf0,cl,cm);
 C[1]= _mm512_mask_xor_epi64(ch,0x0f,ch,cm);	
}
\end{lstlisting}
\normalsize

This ends the computation, the final lines stores the result: the 512 least significant bits in the memory place \texttt{C[0]}, and the most significant bits in \texttt{C[1]}.

\section{Toom-Cook multiplication general algorithm}
\label{app:TCGAlgo}

Several approaches to multiply two arbitrary polynomials over $\mathbb F_2[X]$ of degree at most $N-1$, using the Toom-Cook algorithm, have been presented by Bodrato in \cite{Bodrato07}, Brent \emph{et al.} in \cite{BrentGTZ08}, and software implementations have been provided by Quercia and Zimmermann, in the context of the \texttt{ntl} and the \texttt{gf2x} library, see \cite{Zimmermann08} and \cite{QuerciaZ03}. Let $A$ and $B$ be two binary polynomials of degree at most $N-1$. These polynomials are packed into an array of 64-bit words, whose size is $\lceil N/64\rceil$. Let $t = 3n$ with $n$ a value ensuring $t \geqslant \lceil N/64\rceil$. Now, $A$ and $B$ are considered as polynomials of degree at most $64\cdot t-1$. We discuss the value of $n$ in section \ref{sssec:impliss}.

$A$ and $B$ are split in three parts. One wants now to evaluate the result $C = A\cdot B$ with

$$A = a_0 + a_1\cdot X^{64n} + a_2\cdot X^{2\cdot 64n} \in \mathbb F_2[X],$$
$$B = b_0 + b_1\cdot X^{64n} + b_2\cdot X^{2\cdot 64n} \in \mathbb F_2[X],$$
(of maximum degree $64t-1$, and $a_i, b_i$~of maximum degree~$64n-1$) and,
 
$$C = c_0 + c_1\cdot X^{64n} + c_2\cdot X^{2\cdot 64n} + c_3\cdot X^{3\cdot 64n} + c_4\cdot X^{4\cdot 64n}$$
of maximum degree $ 6\cdot 64n-2$.

The "word-aligned" version evaluates the polynomial for the values $0$, $1$, $x = X^w$, $x+1 = X^w +1$, $\infty$, $w$ being the word size, typically 64 in modern processors. Furthermore, on Intel processors, one can set $w=256$ to take advantage of the vectorized instruction set \texttt{AVX-AVX2}, and even $w=512$ (\texttt{AVX512} extension), at the cost of a slight operand size reduction.

\noindent For the evaluation phase, one has:
$$\begin{array}{lcl}
C(0)	&=& a_0\cdot b_0	\\
C(1)	&=& (a_0 + a_1 + a_2)\cdot (b_0 + b_1 + b_2)\\
C(x)	&=& (a_0 + a_1\cdot x + a_2\cdot x^2)\cdot (b_0 + b_1\cdot x + b_2\cdot x^2)	\\
C(x+1)	&=& (a_0 + a_1\cdot (x + 1) + a_2\cdot (x^2 + 1))\cdot
\\ & & (b_0 + b_1\cdot (x + 1)+ b_2\cdot (x^2 + 1))\\
C(\infty)&=& a_2\cdot b_2
\end{array}
$$
 \normalsize
The implementation of this phase is straightforward, providing that the multiplication $a_i\cdot b_i$ is either another Toom-Cook or Karatsuba multiplication. Notice that the multiplications by $x$ or $x^2$ are virtually free word shifts.

\noindent For the interpolation phase, one has the following equations:

$$\begin{array}{lcl}
C(0)	&=& c_0 	\\
C(1)	&=& c_0 + c_1 + c_2 + c_3 + c_4	\\
C(x)	&=& c_0 + c_1\cdot x + c_2\cdot x^2 + c_3\cdot x^3 + c_4\cdot x^4	\\
C(x+1)	&=& c_0 + c_1\cdot (x + 1) + c_2\cdot (x^2 + 1)\\&& + c_3\cdot (x^3 + x^2 + x + 1) + c_4\cdot (x^4 + 1)	\\
C(\infty)&=& c_4
\end{array}
$$
\normalsize
The matrix associated to this system of equations is given by:

$$M = \left(\begin{array}{ccccc}
                1&                 0&                 0&                 0&                 0\\
                1&                 1&                 1&                 1&                 1\\
                1&                 x&               x^2&               x^3&               x^4\\
                1&             x + 1&           x^2 + 1& x^3 + x^2 + x + 1&           x^4 + 1\\
                0&                 0&                 0&                 0&                 1
\end{array}\right)
$$
and one has :
$$M^{-1} = \left(\begin{array}{ccccc}
                      1&                       0&                       0&                       0&                       0\\
\frac{(x^2 + x + 1)}{(x^2 + x)}&                       1&                     1/x&               \frac 1{x + 1}&                 x^2 + x\\
                      0&             \frac 1{x^2 + x}&              \frac 1{x + 1} &                     1/x&             x^2 + x + 1\\
            \frac 1{x^2 + x}&             \frac 1{x^2 + x}&             \frac 1{x^2 + x}&            \frac 1{x^2 + x} &                       0\\
                      0&                       0&                       0&                       0&                       1
\end{array}\right)
$$

\normalsize

\noindent Finally, the interpolation phase gives :
$$\begin{array}{lcl}
c_0 &=& C(0)\\
c_1 &=& (x^2 + x + 1)/(x^2 + x)\cdot C(0) + C(1) + C(x)/x \\&& + C(x+1)/(x + 1)  + (x^2 + x)\cdot C(\infty)\\
c_2 &=& C(1)/(x^2 + x) + C(x)/(x + 1) + C(x+1)/x \\&&  + (x^2 + x + 1)\cdot C(\infty)\\
c_3 &=&  C(0)/(x^2 + x) + C(1)/(x^2 + x) + C(x)/(x^2 + x) \\&& + C(x + 1)/(x^2 + x)\\
c_4 &=& C(\infty)
\end{array}
$$
\normalsize


\section{Experimentation procedure}
\label{app:perfProc}

Measurements were perfor\-med on a 
Dell Inspiron laptop with an Intel \texttt{Tiger Lake} processor.
\begin{verbatim}
vendor_id	: GenuineIntel
cpu family	: 6
model		: 140
model name	: 11th Gen Intel(R) Core(TM) 
             i7-1165G7 @ 2.80GHz
\end{verbatim}
\normalsize
The compiler is \texttt{gcc} version 10.2.0, the compiler options are as follows:

\noindent\texttt{-O3 -g -march=tigerlake -funroll-all-loops -lm -lgf2x}.

\noindent We kept the \texttt{-funroll-all-loops} option though it does not provide significant improvements. We follow the same kind of test procedure that the one described in \cite{NDruckerGK18}~:
\begin {itemize}
	\item the \textit{Turbo-Boost}\textregistered~is deactivated during the tests;
	\item 1000 runs are executed in order to "heat" the cache memory;
	\item one generates 50 random data sets, and for each data set
      the minimum of the execution clock cycle numbers over a batch of 1000 runs is recorded;
	\item the performance is the average of all these minimums;
	\item this procedure is run on console mode, to avoid system perturbations, and obtain the most accurate cycle counts.
\end{itemize}

The clock cycle counter is \texttt{rdtsc} and the instruction counter is \texttt{rdpmc} with the corresponding selection.
The results for the smallest sizes (i.e. 256 bit and 512 bit operand sizes) are not very reliable since \texttt{rdtsc} and \texttt{rdpmc} are not serializing instructions (see \cite{IntelSD21}). For such sort of small functions, we wanted to avoid the insertion of a costly serializing instruction as \texttt{cpuid}, while the instruction count and the clock cycle number may be less than 20. We chose not to present them. The first size considered is 1024 bits, i.e. binary polynomial of degree at most 1023 operands.

\section{Instruction count and performances}
\label{app:perfProc256}

\subsection{Instruction count comparison}

In Table \ref{tab:instrcount}, we  provide the comparison between the instruction count of our schoolbook and Karatsuba versions. Moreover, we compare this two approaches with the current state-of-the-art \texttt{AVX2} reference. Such an \texttt{AVX2} implementation can be found in the source code of the optimized version of HQC \cite{hqc20}. It uses the \texttt{AVX2} instruction set and the non vectorized \texttt{PCLMULQDQ} instruction. Finally, we also put in Table \ref{tab:instrcount} the instruction number of the assembly source code for the same multiplication presented by Drucker \emph{et al.} in \cite{NDruckerGK18}. Here are some comments on these results:

\begin{itemize}
	\item The best version is our implementation of the schoolbook approach, dividing by more than 2 the instruction number in comparison with the state-of-the-art \texttt{AVX2} implementation.
	\item Our Karatsuba approach presents more instructions but only 3 \texttt{VPCLMULQDQ} instead of 4 for the schoolbook version. Thus, the performance comparison may vary according to the latency and throughput of the instructions.
	\item Drucker \emph{et al.}'s version has 8 \texttt{VPCLMULQDQ} instructions and a larger instruction number (31, instead of 19 for our implementation of the schoolbook approach). This is due to the fact that they only use 2 elementary 64 bit multiplications per \texttt{VPCLMULQDQ} instruction (\texttt{ymmm} version of the instruction), while we use 4. This also implies more XOR's in their case.
\end{itemize}

\subsection{Performances for the 256 bit level kernels}

We present Table \ref{tab:perfKaratRec256} the performances of the \texttt{AVX512} Karatsuba multiplications using the 256 bit kernels presented above. We also include the results of the multiplications using our \texttt{8x8 SB-512} kernel.

\begin{table*}[htbp]
\begin{center}
\begin{tabular}{|c|c|c|c|c|}
\hline
\begin{tabular}{c}Instruction count	\\
	 256 bit size operands
\end{tabular}	&	\multicolumn{3}{c|}{\texttt{VPCLMULQDQ}}	&	\texttt{AVX2}\\
\hline
Instructions						&	SB version	&	Karat. version	& Drucker \emph{et al.} \cite{NDruckerGK18}	&	Karat. Rec.\\
\hline
\texttt{\_mm512\_clmulepi64\_epi128}			&	4	&	3	&	8	&	\\
\hline
\texttt{\_mm\_clmulepi64\_si128}				&		&		&		&	9\\
\hline
\texttt{XOR}									&	5	&	7	&	15	&	25\\
\hline
\texttt{AND}									&	0	&	1	&		&\\
\hline	
\texttt{\_mm512\_broadcast\_i64x4}				&	2	&	2	&		&\\
\hline
\texttt{\_mm512\_permutexvar\_epi64}			&	3	&	6	&	6	&\\
\hline
\texttt{\_mm512\_permutex2var\_epi64}			&	5	&	5	&		&\\
\hline
\texttt{\_mm512\_alignr\_epi64}					&		&		&	2	&\\
\hline
\texttt{\_mm\_loadu\_si128}						&		&		&		&	4\\
\hline
\texttt{\_mm\_shuffle\_epi32}					&		&		&		&	6\\
\hline
\texttt{\_mm\_setzero\_si128}					&		&		&		&	6\\
\hline
\texttt{\_mm\_unpacklo\_epi64}					&		&		&		&	3\\
\hline
\texttt{\_mm\_unpackhi\_epi64}					&		&		&		&	3\\
\hline
\hline
\bf \large Total						&\bf \large 	19	&\bf \large 	24	&\bf \large 	31	&\bf \large 	56	\\
\hline
\hline
additional \texttt{vmovdqa64}'s					&	8	&	9	&	6	&	8\\
\hline
\end{tabular}
\caption{Instruction count for the 256 bit multiplication versions}\label{tab:instrcount}
\end{center}
\end{table*}

\begin{table*}[htbp]
\begin{center}
\begin{tabular}{|c|c|c|c|c|c||c|c|}
\hline
\multicolumn{3}{|c|}{KaratRec}	&	\multicolumn{2}{c|}{Drucker \emph{et al.}}	&	\texttt{AVX2}	&	\multicolumn{2}{c|}{\bf This work}\\
\multicolumn{3}{|c|}{}	&	\multicolumn{2}{c|}{\cite{NDruckerGK18}}	&	\cite{hqc20}	&	\multicolumn{2}{c|}{\texttt{vpclmulqdq-512}}\\
\hline
size	&			&	\texttt{gf2x}	&	\texttt{4x4 - 256}    & \texttt{8x8 - 512}&&	SB-256	&	Karat.-256  \\
\hline
\hline
\multirow{2}{*}{1024}	&

\# clock cycles			&	339		&   239 &   169&	183	&	167	&	186 \\
\cline{2-8}
&\# instructions		&	1224	&	326 &   221&   612	&	276	&	339 \\
\hline
\hline
\multirow{2}{*}{2048}	&

\# clock cycles			&	998	&	744     &   532 &   610	&	541	&	627 \\
\cline{2-8}
&\# instructions		&	3892	&	1137    &   736&   1867	&	908	&	1102 \\
\hline
\hline
\multirow{2}{*}{4096}	&

\# clock cycles			&	2949	&	2262    &   1621&   1929	&	1685	&	1937    \\
\cline{2-8}
&\# instructions		&	11079	&	3632    &   2358&   5684	&	2769	&	3346    \\
\hline
\hline
\multirow{2}{*}{8192}	&

\# clock cycles			&	8742	&	6960    &   4926       &   6038	&	5204	&	6205 \\
\cline{2-8}
&\# instructions		&	33182	&	11535   &   7547    &   17991	&	8718	&	10552\\
\hline
\hline
\multirow{2}{*}{16384}	&

\# clock cycles			&	26128	&	21154   &   14940   &   18327	&	15675	&	18043\\
\cline{2-8}
&\# instructions		&	100163	&	35483   &   23313   &   54840	&	26482	&	32068\\
\hline
\hline
\multirow{2}{*}{32768}	&

\# clock cycles			&	78889	&	65758   &   45244   &   59613	&	47592	&	54881\\
\cline{2-8}
&\# instructions		&	295755	&	108329   &   70941  &   166410	&	80736	&	97278\\
\hline
\hline
\multirow{2}{*}{65536}	&

\# clock cycles			&	226640	&	203855  &   140347  &   187305	&	147572	&	169844\\
\cline{2-8}
&\# instructions		&	853977	&	328772  &   214955  &   503014	&	244222	&	293873\\
\hline
\hline
\multirow{2}{*}{131072}	&

\# clock cycles			&	667900	&	621942  &   425430  &   572984	&	446811	&	511939\\
\cline{2-8}
&\# instructions		&	2516857	&	992919  &   648533  &   1515625	&	736345	&	885164\\
\hline
\end{tabular}
\caption{Performance comparison for Algorithm \ref{algo:karatrec}}\label{tab:perfKaratRec256}
\end{center}
\end{table*}

\end{appendix}

\end{document}